\documentclass[twocolumn, 
 reprint,
superscriptaddress,
 amsmath,amssymb,
 aps,
 pre,
]{revtex4-1}

\usepackage{graphicx}
\usepackage{dcolumn}
\usepackage{bm}
\usepackage{subcaption}
\usepackage{graphicx}
\usepackage[export]{adjustbox}
\usepackage{float}
\usepackage{setspace} 
\usepackage{mathtools}
\usepackage[font=scriptsize,labelfont=bf]{caption}
\usepackage{booktabs,caption}
\usepackage{threeparttable}
\usepackage{framed}

\renewcommand\sectionmark[1]{} 
\makeatletter
\renewcommand\subsectionmark[1]{}
\makeatother

\usepackage[utf8]{inputenc}

\usepackage[dvipsnames]{xcolor}

\begin{document}

\preprint{APS/123-QED}

\title{Path-dependent Dynamics Induced by Rewiring Networks of Inertial Oscillators}

\author{William Qian}%
\affiliation{\mbox{Department of Physics \& Astronomy, College of Arts \& Sciences, University of Pennsylvania, Philadelphia, PA 19104 USA}}

\author{Lia Papadopoulos}
\affiliation{\mbox{Department of Physics \& Astronomy, College of Arts \& Sciences, University of Pennsylvania, Philadelphia, PA 19104 USA}}

\author{Zhixin Lu}
\affiliation{\mbox{Department of Bioengineering, School of Engineering \& Applied Science, University of Pennsylvania, Philadelphia, PA 19104 USA}}

\author{Keith Wiley}
\affiliation{\mbox{Department of Physics \& Astronomy, College of Arts \& Sciences, University of Pennsylvania, Philadelphia, PA 19104 USA}}

\author{Fabio Pasqualetti}
\affiliation{\mbox{Department of Mechanical Engineering, University of California, Riverside, CA 92521 USA}}

\author{Danielle S. Bassett}
\affiliation{\mbox{Department of Physics \& Astronomy, College of Arts \& Sciences, University of Pennsylvania, Philadelphia, PA 19104 USA}}
\affiliation{\mbox{Department of Bioengineering, School of Engineering \& Applied Science, University of Pennsylvania, Philadelphia, PA 19104 USA}}
\affiliation{\mbox{Department of Mechanical Engineering, University of California, Riverside, CA 92521 USA}}
\affiliation{\mbox{Department of Neurology, Perelman School of Medicine, University of Pennsylvania, Philadelphia, PA 19104 USA}}
\affiliation{\mbox{Department of Psychiatry, Perelman School of Medicine, University of Pennsylvania, Philadelphia, PA 19104 USA}}
\affiliation{\mbox{Santa Fe Institute, Santa Fe, NM 87501 USA}}
\affiliation{\mbox{To whom correspondence should be addressed: dsb@seas.upenn.edu}}

\date{\today}

\begin{abstract}
     In networks of coupled oscillators, it is of interest to understand how interaction topology affects synchronization. Many studies have gained key insights into this question by studying the classic Kuramoto oscillator model on static networks. However, new questions arise when network structure is time-varying or when the oscillator system is multistable, the latter of which can occur when an inertial term is added to the Kuramoto model. While the consequences of evolving topology and multistability on collective behavior have been examined separately, real-world systems such as gene regulatory networks and the brain can exhibit these properties simultaneously. How does the rewiring of network connectivity affect synchronization in systems with multistability, where different paths of network evolution may differentially impact system dynamics? To address this question, we study the effects of time-evolving network topology on coupled Kuramoto oscillators with inertia. We show that hysteretic synchronization behavior occurs when the network density of coupled inertial oscillators is slowly varied as the dynamics evolve. Moreover, we find that certain fixed-density rewiring schemes induce significant changes to the level of global synchrony, and that these changes remain after the network returns to its initial configuration and are robust to a wide range of network perturbations. Our findings suggest that the specific progression of network topology, in addition to its initial or final static structure, can play a considerable role in modulating the collective behavior of systems evolving on complex networks.
\end{abstract}

\maketitle


\section{Introduction}
Understanding the emergence of collective behaviors in systems of dynamical units coupled through complex networks remains an important goal in the study of dynamical systems. The synchronization of coupled oscillators is a key example of such behavior \cite{arenas2008synchronization}, and computational models have proven effective in gaining insight into a number of real-world systems where this phenomena occurs, including the synchronization of power grids, the flashing of fireflies, and the dynamics of neuronal networks \cite{Sarfati2020.03.19.999227, motter_myers_anghel_nishikawa_2013, buck_1988, cumin_unsworth_2007, Noori_Park_Griffiths_Bells_Frankland_Mabbott_Lefebvre_2020}. More generally, a number of past studies have focused on the question of how distinct behaviors of coupled oscillators arise from distinct network topologies, assuming that a given topology remains fixed for a given system \cite{Rodrigues_Peron_Ji_Kurths_2016, skardal_2014, Skardal_Restrepo_2012}. Yet, in many systems, network organization is not static, but rather evolves over time; social networks, neuronal networks, and biological regulatory networks are all examples of systems whose interaction topology can change with time \cite{ Noori_Park_Griffiths_Bells_Frankland_Mabbott_Lefebvre_2020,Laurent_Saramaki_Karsai_2015, Calhoun_Miller_Pearlson_Adali_2014, Lebre_Becq_Devaux_Stumpf_Lelandais_2010}. The existence of such time-evolving network systems motivates an investigation of how specific pathways of network evolution alter the dynamical behaviors of coupled oscillators. 

To date, studies of the evolution of coupled oscillators on temporally-evolving networks have often used the Kuramoto model. This model is an established system for studying synchronization behavior, widely used for its simplicity and analytical tractability. For Kuramoto oscillators in the limit of fast network rewiring, prior work has shown that switching between different coupling topologies has the same effect as allowing oscillator dynamics to evolve on a network with weights averaged over the different switching topologies \cite{faggian_2019}. In contrast, another study investigated the effects of network connectivity that co-evolves with Kuramoto oscillator dynamics, and showed that an adaptive rewiring scheme where oscillators re-route links away from their neighbors with which they are most in-phase can result in network topologies that enhance synchronization \cite{papadopoulos_2017}. Other work has demonstrated that networks of phase-lagged Kuramoto oscillators with a biologically-inspired Hebbian learning rule gives rise to unique spatiotemporal modes of oscillation \cite{Timms_English_2014}. These studies are illustrative of the breadth of the field \cite{Yuan_Zhou_2011, Yuan_Zhou_Li_Chen_Wang_2013, Zhou_Kurths_2006, Aoki_Aoyagi_2009, Aoki_Aoyagi_2011, Baldi_Tao_Kosmatopoulos_2019}, which collectively demonstrates that the dynamics of Kuramoto oscillators depend appreciably on the type of reconfiguration that the coupling network undergoes.

In the presence of a bimodal natural frequency distribution, phase lags, or frequency-degree correlations, the Kuramoto model can exhibit a variety of complex behaviors, such as hysteretic transitions as a function of coupling strength \cite{pazo_montbrio_2009, metivier_gupta_2019, coutinho_goltsev_dorogovtsev_mendes_2013, Yeung_Strogatz_1999}. However, under the most generic conditions, the Kuramoto model does not exhibit path-dependent dynamics. Specifically, when adiabatically increasing and then decreasing the coupling strength of a Kuramoto oscillator population, the value of the order parameter is typically identical along the forward and backward transitions. Indeed, for non-negative values of coupling, systems of Kuramoto oscillators are monostable \cite{Esmaeili_2017, Labavic_Meyer-Ortmanns_2017}, indicating that the path of network evolution taken during oscillator dynamics does not affect levels of synchrony once a coupling pattern has been fixed. In other words, history has no effect on the dynamics of standard Kuramoto oscillators once transient effects are discarded. 

In contrast, systems of second-order Kuramoto oscillators are known to be sensitive to history. In particular, the introduction of an inertial term to the Kuramoto model has been shown to result in highly multistable dynamics in certain parameter regimes \cite{olmi_2015, Jaros_Brezetsky_Levchenko_Dudkowski_Kapitaniak_Maistrenko_2018}. Unlike in the standard Kuramoto model, adiabatically tuning the coupling strength of inertial Kuramoto oscillators results in hysteretic synchronization transitions \cite{olmi_navas_boccaletti_torcini_2014}. Specifically, slowly increasing and then decreasing the coupling strength of inertial Kuramoto oscillators creates a hysteresis loop in the order parameter, indicating that inertial oscillator dynamics can depend significantly on prior conditions. Furthermore, it has been analytically proven that any nonzero amount of inertia can induce these hysteresis loops by turning a supercritical bifurcation into a subcritical bifurcation \cite{barre_2016}. These behaviors and the studies unearthing them collectively suggest that path-dependent dynamics may arise from time-varying connectivity in networks of inertial Kuramoto oscillators. 

Like the standard Kuramoto model, the inertial Kuramoto model has also proven insightful for understanding real-world systems. The inertial Kuramoto model was first introduced to explain synchronization patterns in groups of fireflies \cite{tanaka_1997_PRL}. It has since been used extensively to study the stability of power grids and the synchronization of Josephson junctions \cite{wiesenfeld_colet_strogatz_1998}. One interpretation of the inertial term is that it extends the Kuramoto model beyond the simplified and completely overdamped regime, where system dynamics behave analogously to coupled units oscillating in an extremely viscous medium. The inclusion of inertia allows for both underdamped and overdamped dynamics, depending on the value of the inertial constant. In the context of neuroscience, one source of biological support for the inclusion of an inertial term in equations for neural dynamics comes in the form of inertia being analogous to inductance \cite{Cao_Wan_2014}. Forms of inductance have been observed experimentally in squid axons, and the inclusion of inductive effects in models of neurons has been shown to allow for richer modulation of temporal dynamics \cite{mauro_conti_dodge_schor_1970, koch_1984}. Other studies have used inertial phase oscillators as a simplified model for the dynamics of a neuron with an axon and dendrite \cite{dolan2005phase,majtanik2006desynchronization}, finding that incorporating inertia to model dendritic dynamics can alter responses to stimulation.

Given the sensitivity of the inertial Kuramoto model to history, we hypothesize that network rewiring alone can induce path-dependent behavior in networks of inertial Kuramoto oscillators. This hypothesis, combined with the relevance of the inertial Kuramoto model to real-world systems, prompts us to investigate how specific network evolution pathways affect the collective behavior of inertial Kuramoto oscillators. In particular, some important questions arise: how does the process of network evolution impact the dynamics of a population of inertial oscillators? In addition, which paths of network evolution are effective in synchronizing or desynchronizing inertial oscillators, and to what extent do these effects persist after further network rewiring? Previous studies have shown that for standard Kuramoto oscillators with static connectivity, modular network structures promote local synchrony but hinder global synchronization \cite{Skardal_Restrepo_2012, Oh_Rho_Hong_Kahng_2005}. In addition, prior work suggests that high alignment between the eigenvectors of the network Laplacian and the oscillators' natural frequencies can enhance synchronization in the standard Kuramoto model \cite{skardal_2014}. These results prompt an analysis of whether modular networks and synchrony-optimized networks behave similarly for the case of inertial oscillators at each node, and how paths of network evolution towards or away from these special topologies may affect system dynamics. 

The remainder of the paper is organized as follows. Section~\ref{sec:II} introduces the inertial Kuramoto model on complex networks. In Section~\ref{sec:III}, we examine how the collective dynamics of inertial oscillators are affected by different network evolution schemes, and we analyze the robustness of effects induced by network rewiring. We conclude in Section~\ref{sec:IV} with a discussion of our findings as well as of possible areas for further study.

\section{The Inertial Kuramoto Model}
\label{sec:II}
A system of $N$ inertial Kuramoto oscillators evolves according to the equation  
\begin{equation}
m \ddot \theta_i + \dot \theta _i  = \omega _i + \alpha \sum_{j=1}^{N}A_{ij}\sin(\theta _j - \theta _i),    
\end{equation}
where $\theta _i$ represents the instantaneous phase of the $i$th oscillator, $\omega_i$ is the natural frequency of the oscillator, $\alpha$ is the coupling strength, $m$ is the inertial constant, and $\boldsymbol{A}$ is an $N \times N$ unweighted, undirected adjacency matrix representing network connectivity. Note that in the overdamped limit $m \rightarrow 0$, the original first-order Kuramoto model is recovered.

The instantaneous level of global synchrony in a population of oscillators is usually quantified by the modulus of the complex order parameter
\begin{equation}
    R(t) = \frac{1}{N}\left|\sum_{j=1}^{N}e^{i\theta _j (t)}\right|,
\end{equation}
which takes on values ranging from $0$ to $1$, with higher values indicating higher levels of phase synchronization. We also introduce the time-averaged order parameter
\begin{equation}
    \langle R \rangle = \frac{1}{T}\int_{T_R}^{T_R+T} R(t) dt,
\end{equation}
where $T_R$ represents a discarded transient period, and $T$ is the length of the interval over which the order parameter is averaged.

\begin{figure}
    \includegraphics[trim = 1000 500 800 300, scale=.22]{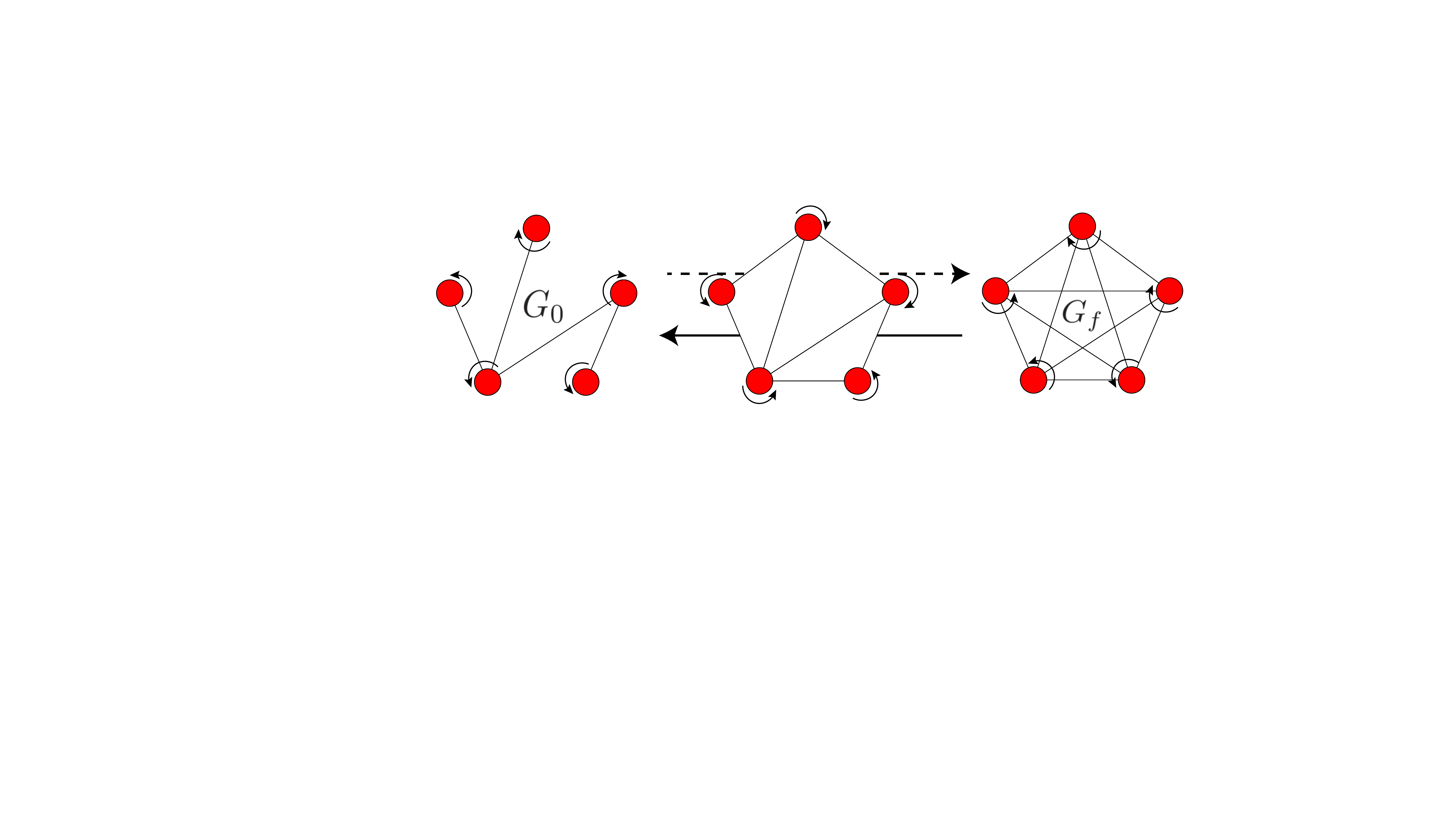}
    \caption{\textbf{A schematic of the network rewiring process.} Initially, the network connectivity evolves from $G_0$ to $G_f$ through a series of unweighted and undirected intermediate graphs (dashed arrow). The network connectivity then returns to $G_0$ through the same series of intermediate graphs (solid arrow). Throughout this rewiring process, oscillator dynamics evolve atop the time-varying connectivity.}
    \label{fig:my_label}
\end{figure}

\section{Simulations and Rewiring Procedures}
\label{sec:III}

We used $N = 100$ oscillators in all simulations. Initial phases $\{\theta _i (0) \}$ were selected at random from $[-\pi, \pi]$, while initial frequencies $\{\dot \theta _i(0) \}$ and natural frequencies $\{\omega _i \}$ were both selected at random from a uniform distribution in the interval $[-3, 3]$. Unless specified otherwise, reported measures represent ensemble averages over different graph structures, initial conditions, and natural frequencies.

To understand how time-varying connectivity affects networked inertial oscillators, we developed a network rewiring scheme that allowed us to isolate the effects of network rewiring on oscillator dynamics. Given initial and final graphs $G_0 = (V, E_0)$ and $G_f = (V, E_f)$, we generated a sequence of intermediate graphs $\{G_0, G_1, \dots G_f\}$ that determined how network topology would vary over time. Specifically, let $S_{del} = E_0 \setminus E_f $ denote the edges in $G_0$ but not in $G_f$, and let $S_{add} = E_f \setminus E_0 $ be the edges in $G_f$ but not in $G_0$. We generate the $i+1^{st}$ intermediate graph $G_{i+1}$ from $G_{i}$ by randomly removing $\approx |S_{del}|/f$ edges in $S_{del} \cap E_{i}$ from $G_i$, and randomly adding $\approx |S_{add}|/f$ edges in $S_{add} \cap \overline{E_{i}}$ to $G_i$, where $\overline{G} = (V, \overline{E})$ represents the complement graph of $G$. 

After generating the sequence of graphs $\{ G_0, \dots , G_f \}$, we carried out a two-step process (Fig.~\ref{fig:my_label}). First, we simulated the time-evolution of inertial oscillator dynamics as network connectivity evolved from $G_0$ to $G_f$ through the series of intermediate networks. Then, we continued the time-evolution of inertial oscillator dynamics as network connectivity evolved from $G_f$ back towards $G_0$ through the same series of intermediate graphs. For our simulations, we use $f = 50$ transition graphs, and the network rewiring occurs every $l = 5 \times 10 ^ 4$ time-steps at $\Delta t = 0.02$ resolution. We also confirm that our main results hold when using different timescales of rewiring, as well as when using a different number of transition graphs (see Supplementary Figs.~\ref{fig:10} and~\ref{fig:11}). Time-averaged values of the order parameter at each network in the rewiring process are reported after discarding a transient period of $T_{R} = \frac{1}{2} (l \times \Delta t)$. Additional initial and final transients of simulations are also explicitly shown where appropriate. 

\begin{figure*}[!htb]
\begin{subfigure}[t]{0.03\textwidth}
  \vspace{-5.2cm}
  (a)
\end{subfigure}
  \adjustbox{minipage=1.3em,valign=t}{\label{sfig:testa}}%
  \begin{subfigure}[t]{\dimexpr.485\linewidth-1.3em\relax}
  \centering
  \includegraphics[trim = 1210 150 1000 200, scale=.18]{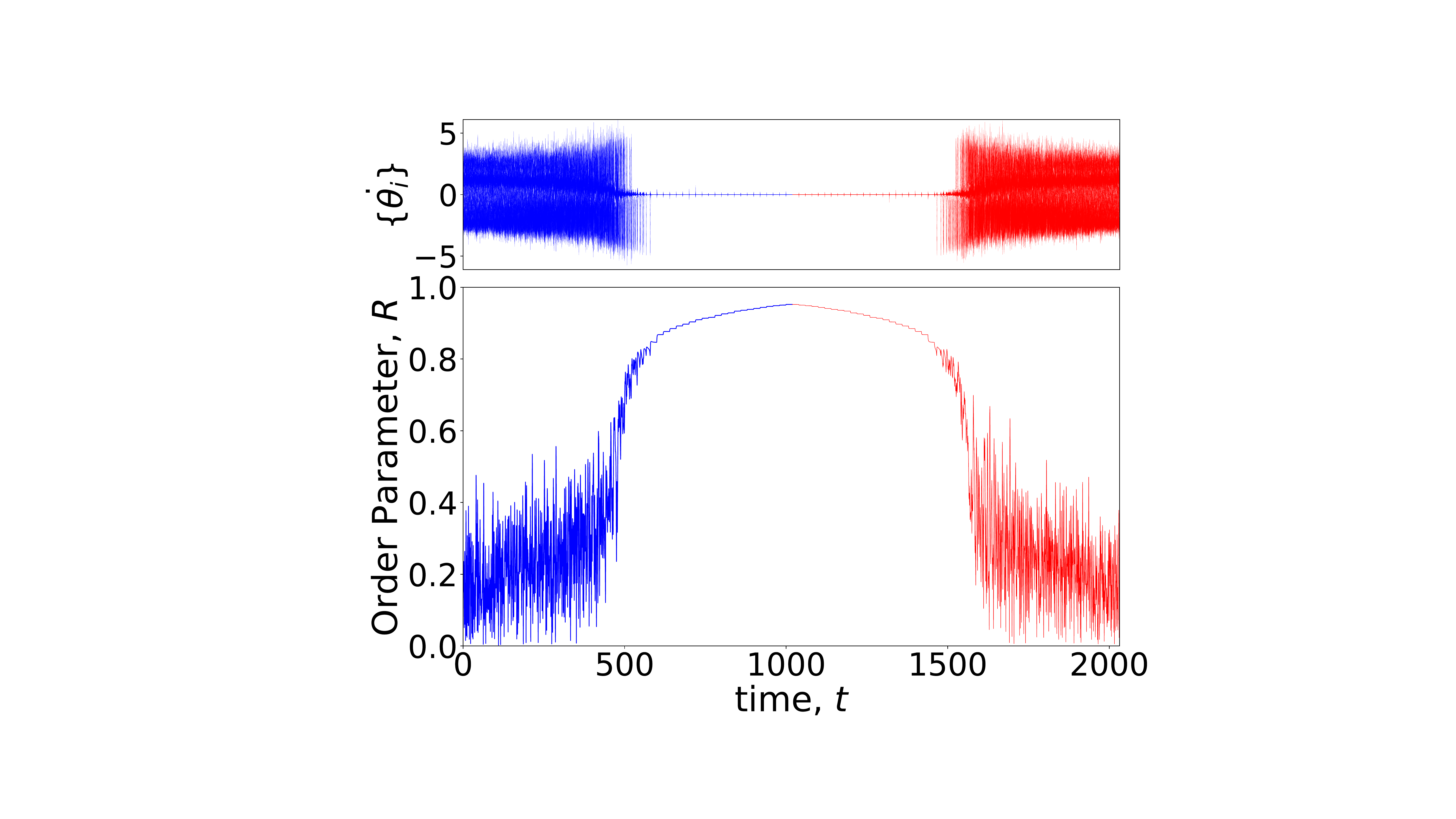}
  \end{subfigure}%
  \begin{subfigure}[t]{0.03\textwidth}
  \vspace{-5.2cm}
  (b)
\end{subfigure}
  \adjustbox{minipage=1.3em,valign=t}{\label{sfig:testb}}%
  \begin{subfigure}[t]{\dimexpr.485\linewidth-1.3em\relax}
  \includegraphics[trim = 1210 150 1000 200, scale=.18]{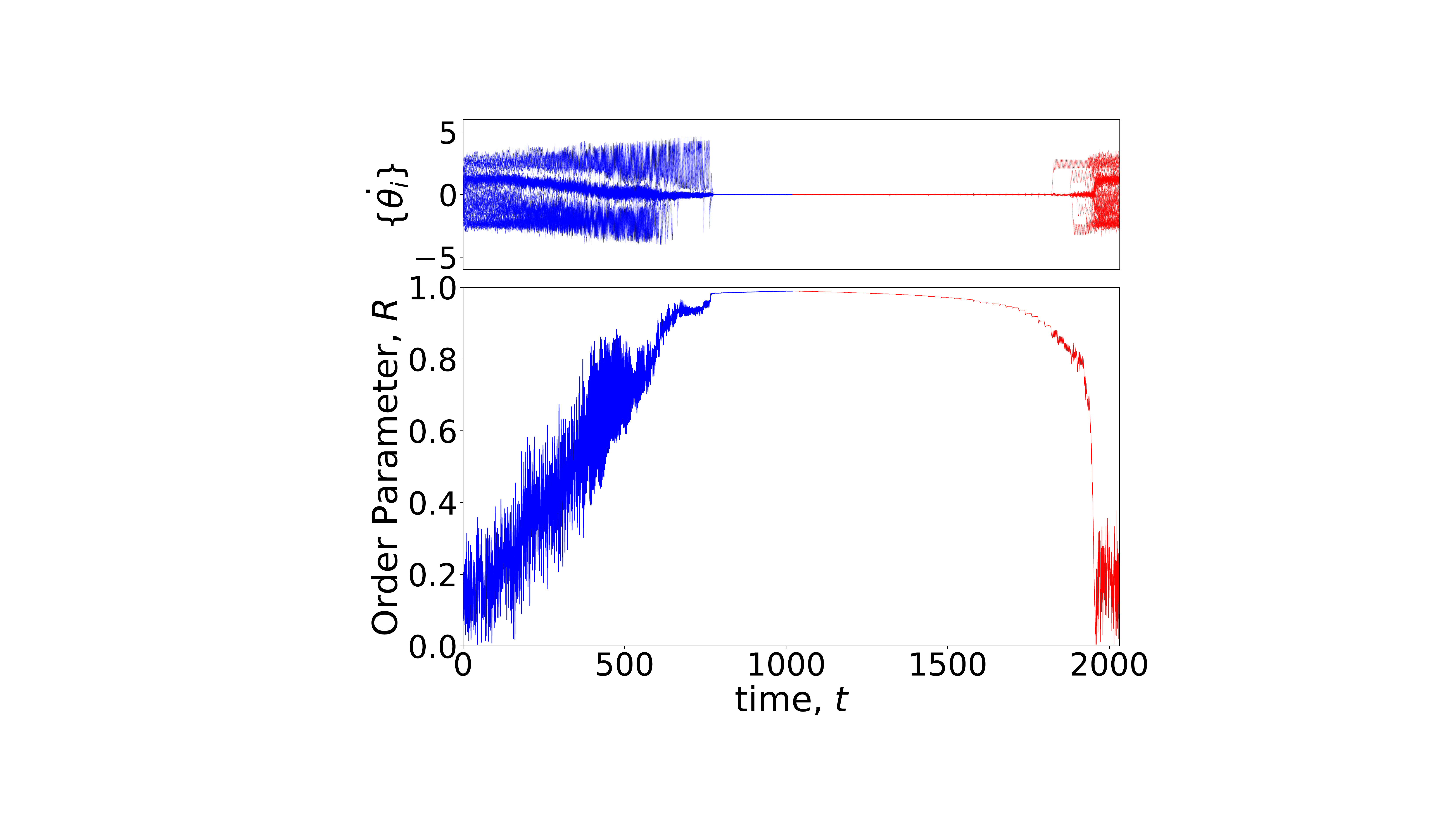}
  \end{subfigure}
  \caption{\textbf{Oscillator dynamics when varying network density, with and without inertia.} Single-instance examples of all oscillators' instantaneous frequencies $\{ \dot{\theta_i} \}$ (top panels) and the global order parameter $R(t)$ (bottom panels) as a function of time as network density is first increased ($\langle k \rangle = 10 \rightarrow 40$) and then decreased ($\langle k \rangle = 40 \rightarrow 10$) as the dynamics run. \emph{(a)} The process with no inertia ($m = 0$, $\alpha = 0.15$). \emph{(b)} The process with inertia ($m = 2$, $\alpha = 0.3$). Network evolution from $G_0$ to $G_f$ and from $G_f$ back to $G_0$ are colored blue and red, respectively. For visual clarity, these examples were produced on a faster rewiring timescale than described in the main text ($l = 10^3$). Parameters have been chosen such that minimum and maximum levels of synchrony are comparable in the two cases.}
  \label{fig2}
\end{figure*}

\subsection{Varying Network Density}

\begin{figure*}
\begin{subfigure}[t]{0.03\textwidth}
  \vspace{-5.2cm}
  (a)
\end{subfigure}
  \adjustbox{minipage=1.3em,valign=t}{\label{sfig:testa}}%
  \begin{subfigure}[t]{\dimexpr.485\linewidth-1.3em\relax}
  \centering
  \includegraphics[trim = 1000 250 1000 250, scale=.24]{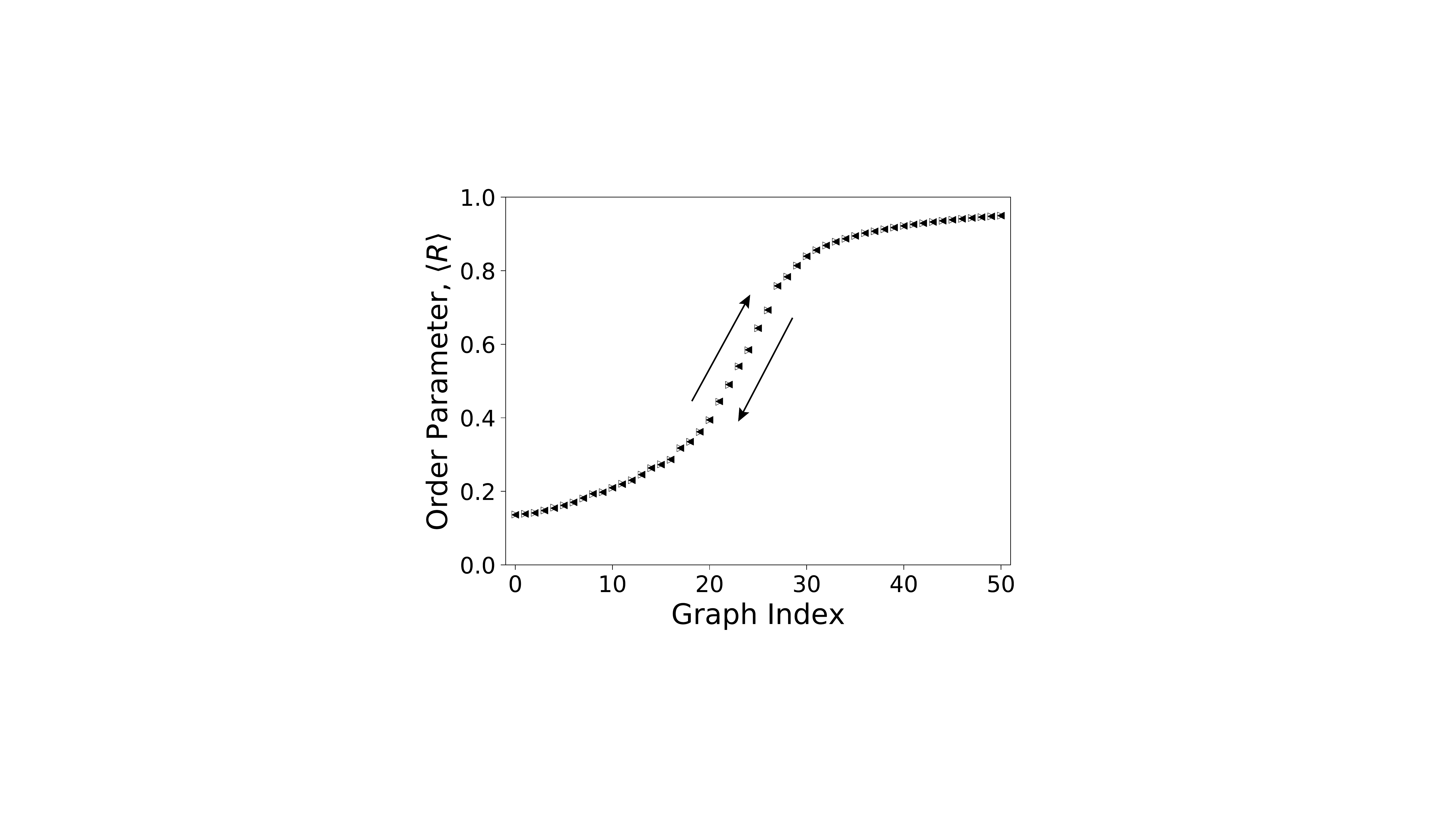}
  \end{subfigure}%
  \begin{subfigure}[t]{0.03\textwidth}
  \vspace{-5.2cm}
  (b)
\end{subfigure}
  \adjustbox{minipage=1.3em,valign=t}{\label{sfig:testb}}%
  \begin{subfigure}[t]{\dimexpr.485\linewidth-1.3em\relax}
  \centering
  \includegraphics[trim = 1000 250 1000 250, scale=.24]{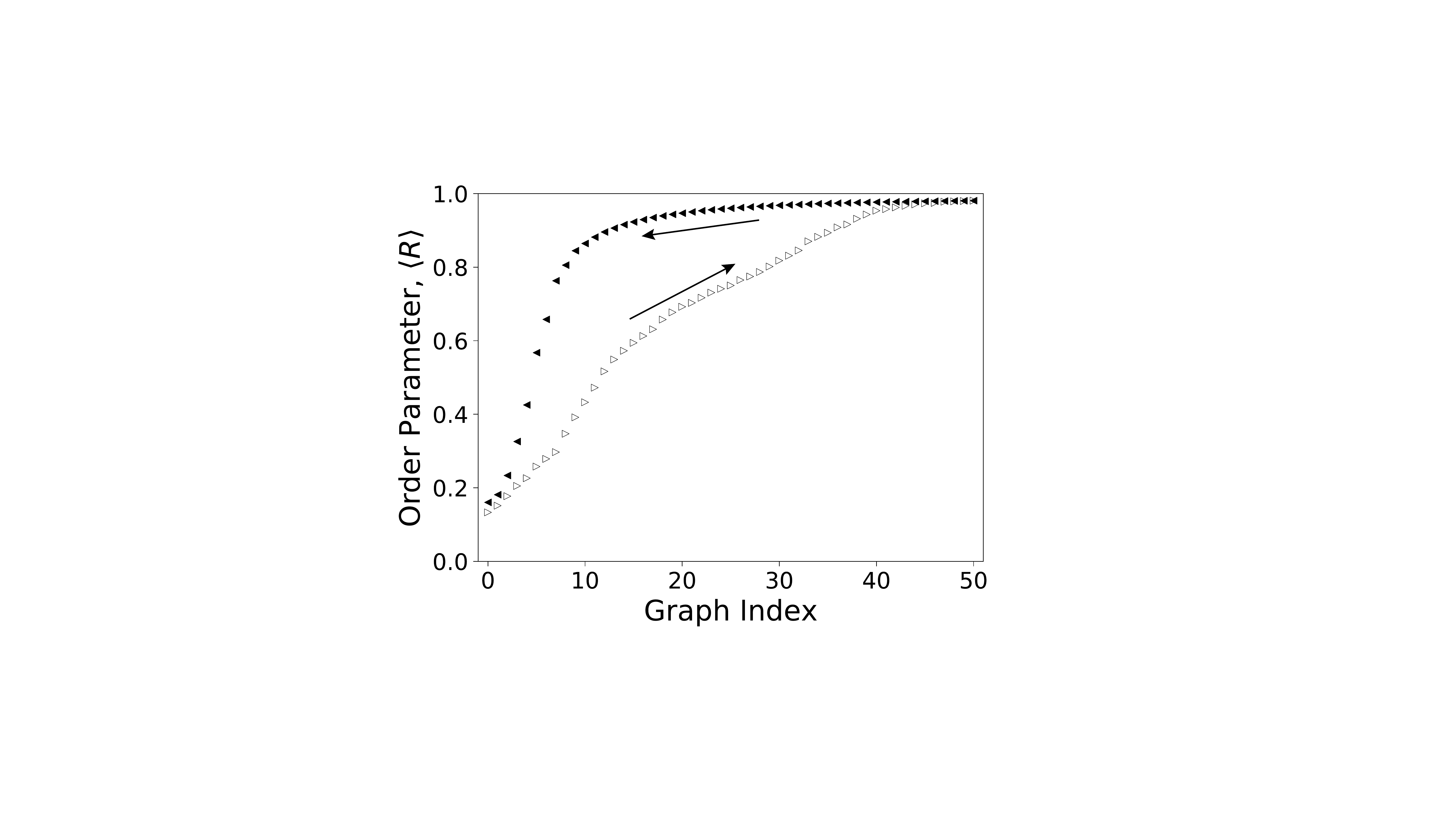}
  \end{subfigure}
  \caption{\textbf{Inertia causes hysteresis in the order parameter when varying network density.} The time-averaged order parameter $\langle R \rangle$ as network density is first increased ($\langle k \rangle = 10 \rightarrow 40$) and then decreased ($\langle k \rangle = 40 \rightarrow 10$). \emph{(a)} The process with no inertia ($m = 0$, $\alpha = 0.15$). \emph{(b)} The process with inertia ($m = 2$, $\alpha = 0.3)$. The coupling strength in panel \emph{(a)} has been chosen to ensure that minimum and maximum levels of synchrony are comparable in the two cases. In both panels, the order parameter is plotted against the graph index of the intermediate graphs (rather than the average degree) to emphasize that $\langle R \rangle$-values lying along the same vertical line were obtained from identical oscillator network-connectivity. All curves depict averages over 25 instantiations.}
    \label{fig:3}
\end{figure*}

We first demonstrate that hysteretic synchronization behavior occurs while increasing and then decreasing network density as oscillator dynamics evolve atop the time-varying network structure. We generate graphs $G_0$ through $G_{f}$ by starting with a random Erdős–Rényi graph with an average degree of $\langle k \rangle = 10$. We next add edges uniformly at random until a graph $G_{f}$ with an average degree of $\langle k \rangle = 40$ is reached. Then, we apply the rewiring procedure (see Sec. ~\ref{sec:III}) to generate all intermediate graphs between $G_0$ and $G_{f}$. Starting with $G_0$ we allow the dynamics of the oscillators to run atop the graph with random initial conditions $\{\theta_i(0)\}$ and $\{\dot \theta_i(0) \}$. Next, we switch the network topology to $G_1$, a slightly denser graph, using the final states of the oscillators from running atop $G_0$ as the initial conditions for $G_1$. This sequential process is repeated until $G_{f}$ is reached, and is then continued in reverse until the network returns to $G_0$. We hold $m$ and $\alpha$ constant throughout the process. 

To determine if the presence of inertia gives rise to path-dependent behavior, we allow Kuramoto oscillator dynamics to evolve with and without inertia while we vary network density in the manner described above. To compare the two situations, we set the coupling values for the non-inertial and inertial system such that the initial level of synchrony is relatively low and comparable between the two cases. As expected, in the absence of inertia, we find that oscillator dynamics evolve in a reversible manner throughout the $G_0 \rightarrow G_f \rightarrow G_0$ rewiring process, suggesting that dynamics are identical for the same network structures regardless of the network evolution pathway taken to reach those structures (Fig.~\ref{fig2}a). However, this reversibility is not observed in the presence of inertia (Fig.~\ref{fig2}b), where we instead observe asymmetric trajectories of both phase synchronization and frequency entrainment as a function of time.

Given the irreversibility of collective dynamics in the inertial case, we hypothesized that a hysteresis loop of the time-averaged order parameter should form as the network density is slowly increased and then decreased back to its initial value. We indeed observe this phenomena when inertia is present (Fig.~\ref{fig:3}b), but not for the standard Kuramoto system (Fig.~\ref{fig:3}a). Note that this finding is consistent with prior work reporting hysteretic behavior in the second-order Kuramoto model while tuning the coupling strength but holding network connectivity fixed \cite{olmi_navas_boccaletti_torcini_2014}. Indeed, for Erdős–Rényi random networks, it is intuitive that increasing and then decreasing network density should have an effect similar to that of increasing and then decreasing the coupling strength. 

\begin{figure*}
\begin{subfigure}[t]{0.03\textwidth}
  \vspace{-5.2cm}
  (a)
\end{subfigure}
  \adjustbox{minipage=1.3em,valign=t}{\label{sfig:testa}}%
  \begin{subfigure}[t]{\dimexpr.485\linewidth-1.3em\relax}
  \centering
     \includegraphics[trim = 1110 200 1100 210, scale=.21]{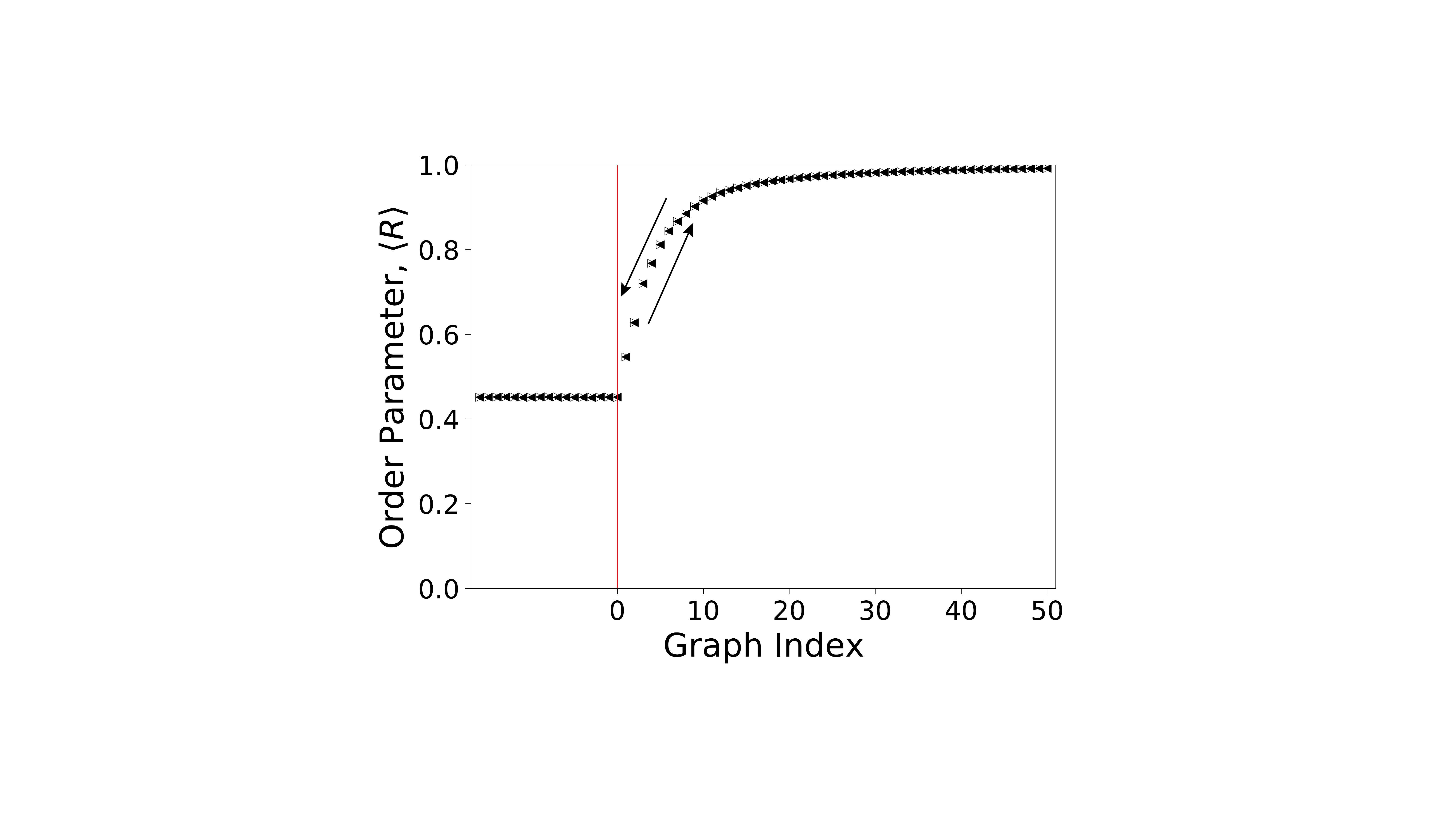}
  \end{subfigure}%
  \begin{subfigure}[t]{0.03\textwidth}
  \vspace{-5.2cm}
  (b)
\end{subfigure}
  \adjustbox{minipage=1.3em,valign=t}{\label{sfig:testb}}%
  \begin{subfigure}[t]{\dimexpr.485\linewidth-1.3em\relax}
  \centering
  \includegraphics[trim = 1110 200 1100 210, scale=.21]{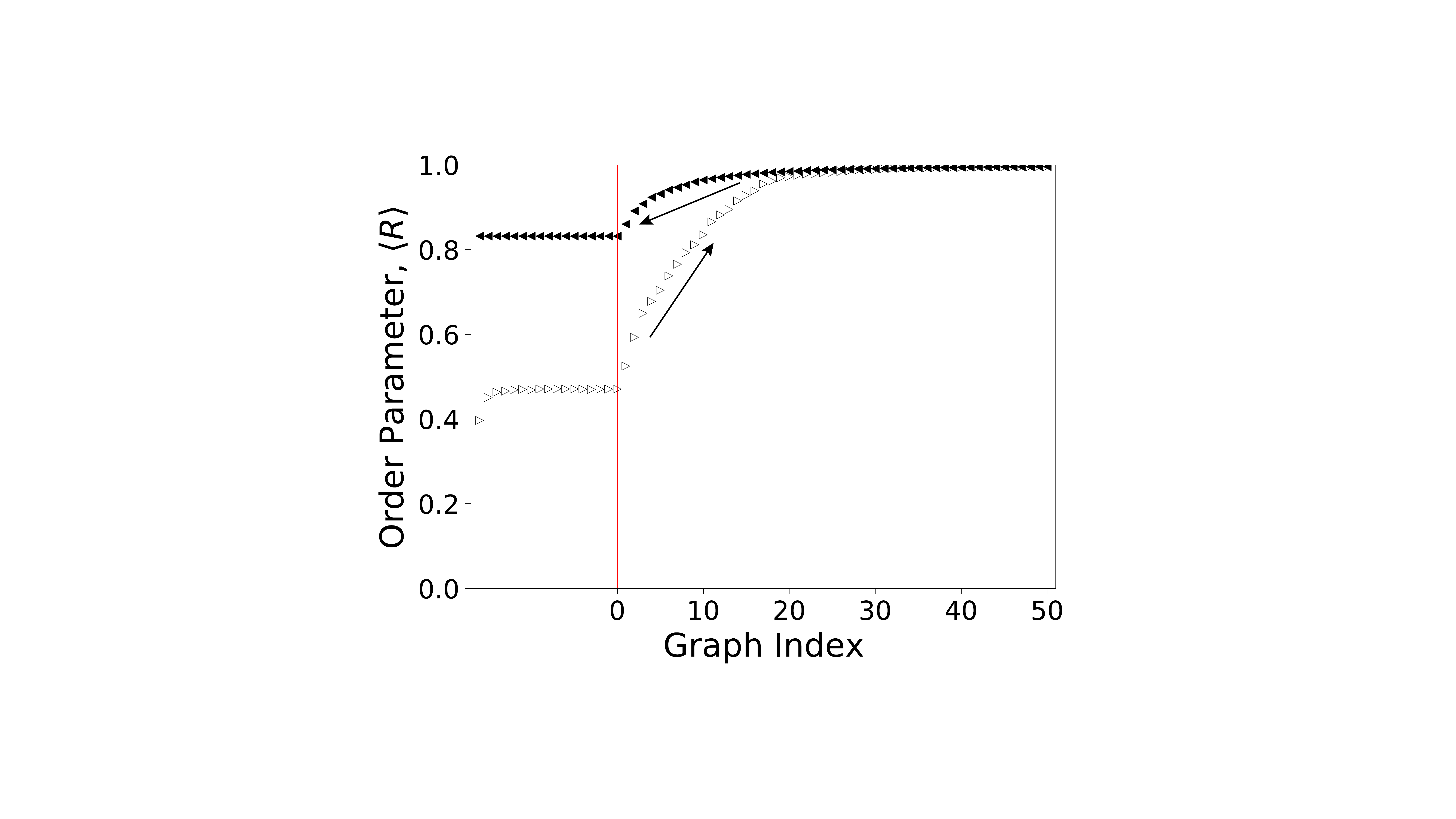}
  \end{subfigure}
      \caption{\textbf{Varying network density with strong coupling.} The time-averaged order parameter $\langle R \rangle$ as network density is first increased ($\langle k \rangle = 10 \rightarrow 40$) and then decreased ($\langle k \rangle = 40 \rightarrow 10$) with strong coupling. \emph{(a)} The process with no inertia ($m = 0$, $\alpha = 0.35$). \emph{(b)} The process with inertia ($m = 2$, $\alpha = 0.5$). Network evolution occurs to the right of the red line, with initial and final transients shown to the left. The coupling strength in panel \emph{(a)} has been chosen to ensure that the minimum and maximum levels of synchrony are comparable in the two cases. All curves depict an average over 25 instantiations.}
  \label{fig:4}
\end{figure*}

Our observations thus far leave unanswered the question of how varying network density in the manner we describe affects oscillator dynamics when both inertia and strong network coupling are present. To investigate this case, we increased the global coupling strength $\alpha$ for both the inertial and non-inertial system such that the initial synchrony level would be intermediately-valued and again approximately the same for the two conditions. In this manner, the oscillators would initially exhibit partially synchronized dynamics in both cases. At high coupling, the order parameter for the model without inertia continues to exhibit reversible behavior (Fig.~\ref{fig:4}a). In contrast, transitions from low-density networks towards and away from high-density networks create a significant separation between the forward and backward order parameter curves when inertia is present (Fig.~\ref{fig:4}). However, the form of the irreversibility at high coupling is qualitatively different than that observed with moderate coupling (Fig.~\ref{fig:3}b). As opposed to the case with moderate coupling, no closed hysteresis loop is formed.
Rather, at high coupling, levels of synchrony remain markedly increased even after the original lowest-density network is recovered. The shape of this trajectory suggests that, when the parameters and initial network connectivity of inertial oscillators allow for partially synchronized oscillator dynamics, network evolution towards and away from more synchronizable network structures may irreversibly increase the levels of global synchrony.  

\subsection{Constant Network Density}

The changes we report in the rewiring processes described above could be a function of both the changing topology and the changing network density. To isolate the effects of changing topology it is therefore of interest to consider rewiring processes that maintain the network density. This case is also particularly relevant to real-world network systems wherein there often exists a cost associated with the development and maintenance of network connections. For example, the energy consumed by synapses in mammalian brains places metabolic constraints on brain development \cite{Karbowski_2012, Fonseca-Azevedo_Herculano-Houzel_2012}. A question then arises whether there exist fixed-density rewiring schemes that also produce significant separation between the forward and backward order parameter curves. To answer this question, it is useful to consider network evolution pathways toward or away from topologies known to be highly synchronizable in the standard Kuramoto model. Prior work has demonstrated that networks of standard, first-order Kuramoto oscillators with optimal alignment between the network Laplacian's eigenvectors and the oscillators' natural frequencies are highly synchronizable \cite{skardal_2014}. To describe this alignment, let $\lambda _j$ and $\boldsymbol{v}^j$ represent the $j$-th largest eigenvalue and its corresponding eigenvector of the network Laplacian $L_{ij} = \delta _{ij} k_i -A_{ij}$, where $k_{i}$ is the degree of node $i$. Following Ref. \cite{skardal_2014}, in the strongly synchronized regime, minimizing the synchrony alignment function 
\begin{equation}
    J(\boldsymbol{\omega}, L) = \frac{1}{N}\sum_{j=2}^{N}\lambda _j ^{-2}\langle \boldsymbol{v}^{j},\boldsymbol{\omega}\rangle ^2,
\end{equation}
serves to maximize the global order parameter $R$ in the standard Kuramoto model.

Applying this approach, we generated synchrony-aligned networks of a given average degree $\langle k \rangle$ via a hill-climbing algorithm with the procedure described in Ref. \cite{skardal_2014} (See Supplementary Material). Then, using Erdős–Rényi graphs for $G_0$ and synchrony-aligned graphs for $G_f$, we simulated the network evolution pathway defined by $G_0 \rightarrow G_f \rightarrow G_0$ while maintaining a fixed network density ($\langle k \rangle = 20$). 

\begin{figure*}
\begin{subfigure}[t]{0.03\textwidth}
  \vspace{-5.6cm}
  (a)
\end{subfigure}
  \adjustbox{minipage=1.3em,valign=t}{\label{sfig:testa}}%
  \begin{subfigure}[t]{\dimexpr.485\linewidth-1.3em\relax}
  \centering
     \includegraphics[trim = 1030 200 1000 210, scale=.22]{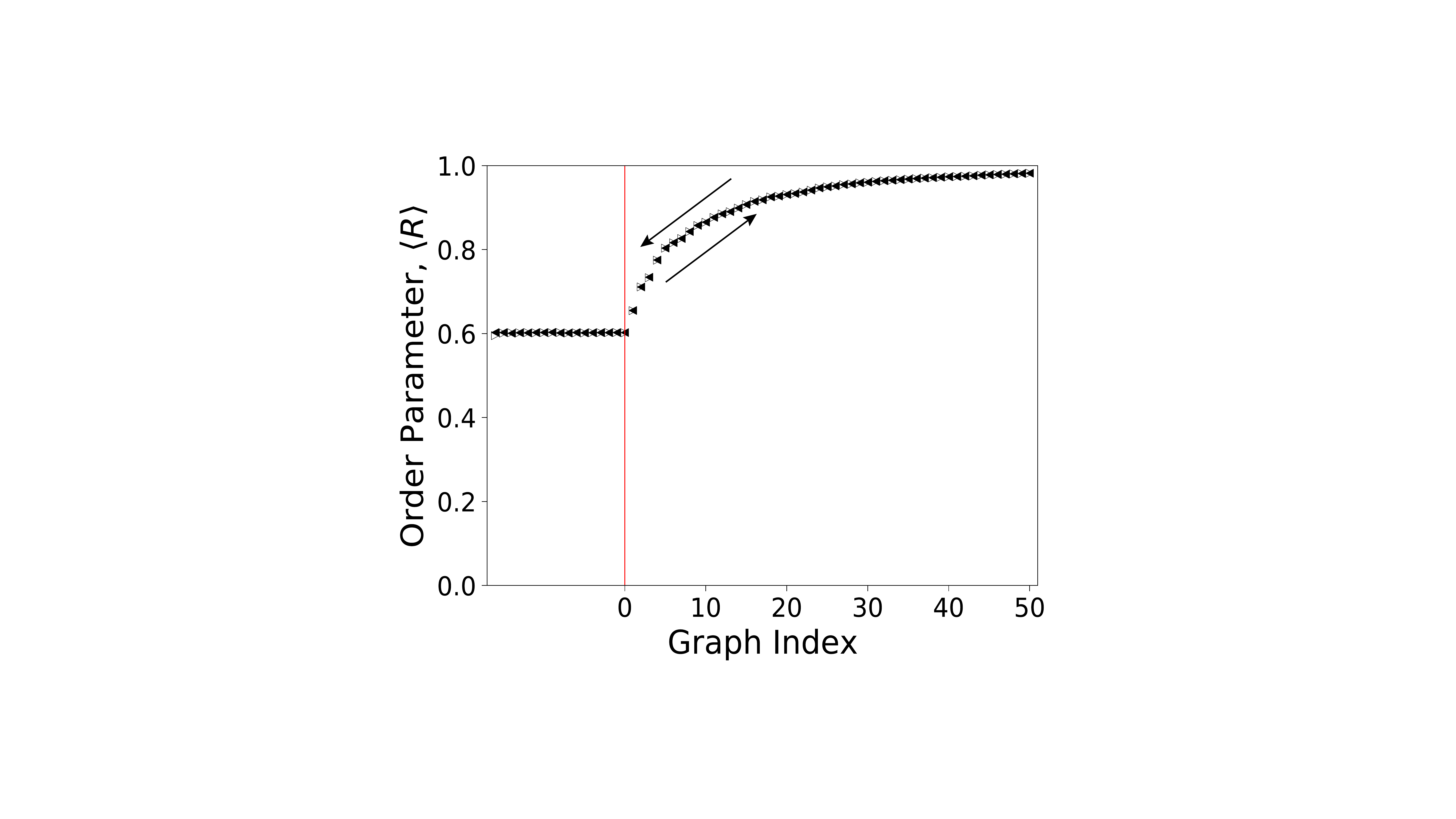}
  \end{subfigure}%
  \begin{subfigure}[t]{0.03\textwidth}
  \vspace{-5.6cm}
  (b)
\end{subfigure}
  \adjustbox{minipage=1.3em,valign=t}{\label{sfig:testb}}%
  \begin{subfigure}[t]{\dimexpr.485\linewidth-1.3em\relax}
  \centering
  \includegraphics[trim = 1220 200 1000 210, scale=.22]{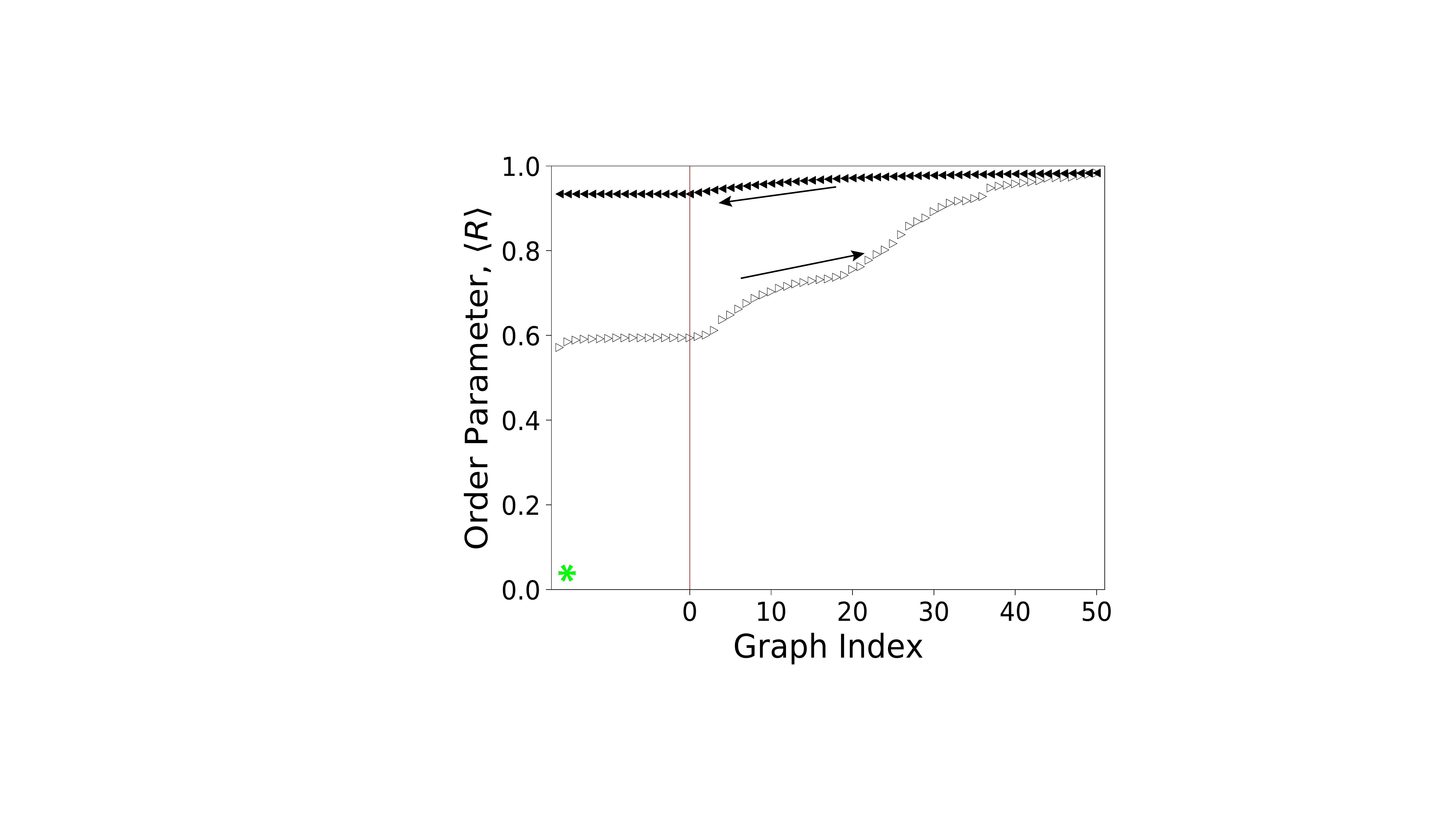}
  \end{subfigure}
      \caption{\textbf{Synchrony gains through network evolution at constant density and intermediate coupling.} The time-averaged order parameter $\langle R \rangle$ as the network connectivity of inertial oscillators evolves from an Erdős–Rényi graph towards a synchrony-aligned graph at constant density ($\langle k \rangle$ = 20), followed by reversal along the same set of intermediate networks until the original connectivity graph is recovered. \emph{(a)} The process with no inertia ($m = 0$, $\alpha = 0.18$). \emph{(b)} The process with inertia ($m = 2$, $\alpha = 0.3$). Note that coupling strengths have been chosen such that minimum and maximum levels of synchrony are comparable in the two cases depicted in panels \emph({a}) and \emph{(b)}. All curves depict averages over 25 instantiations.}
  \label{fig:5}
  \vspace{-.5em}
\end{figure*}

We begin by considering a situation of relatively high global coupling (Fig.~\ref{fig:5}). For the standard Kuramoto model, rewiring towards synchrony-aligned networks increases the order parameter substantially, and as expected, the synchrony level returns to its initial value along the same path as the network returns back to the original Erdős–Rényi graph (Fig.~\ref{fig:5}a). When inertia is incorporated (and the coupling strength adjusted to obtain a similar level of initial synchrony), we again find that network evolution towards synchrony-aligned graphs enhances the order parameter, and the system nears perfect synchrony at $G_f = G^{*}$ (Fig.~\ref{fig:5}b). Moreover, in contrast to the non-inertial case, the transition from Erdős–Rényi graphs toward and away from synchrony-aligned graphs creates a significant separation between the forward and backward order parameter curves. However, this rewiring does not elicit a closed hysteresis loop. Similar to the case of varying network density with strong coupling (Fig.~\ref{fig:4}) for the second-order Kuramoto model, we find that the steady-state level of synchrony is maintained at a significantly higher value even after the system returns to the original Erdős–Rényi graph. 

\begin{figure*}
\begin{subfigure}[t]{0.02\textwidth}
  \vspace{-6cm}
  (a)
\end{subfigure}
  \adjustbox{minipage=1.3em,valign=t}{\label{sfig:testb}}%
  \begin{subfigure}[t]{\dimexpr.485\linewidth-1.3em\relax}
  \centering
  \hspace*{.8cm}\includegraphics[trim = 1400 -100 1100 0, scale=.15]{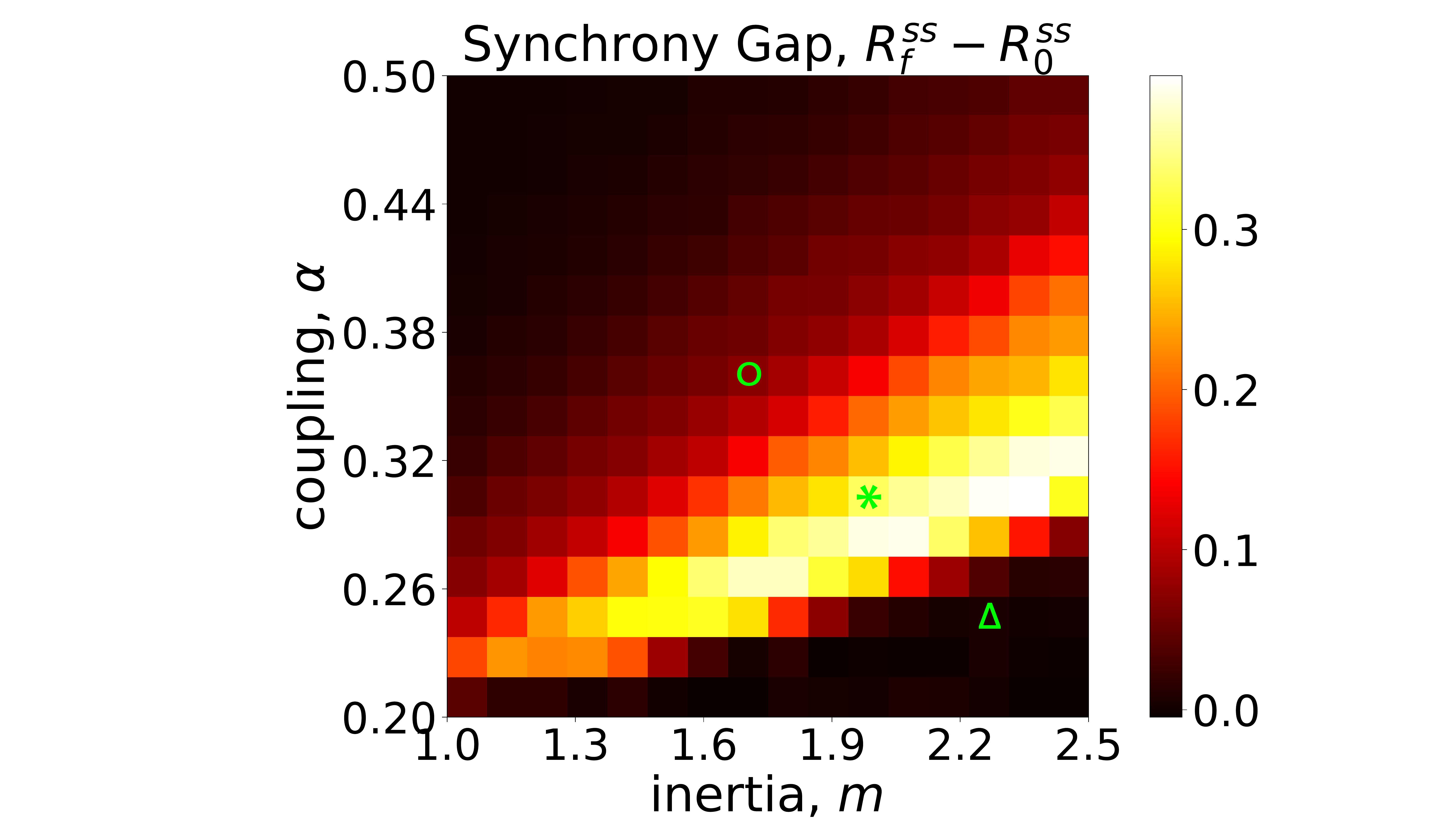}
  \end{subfigure}
  \begin{subfigure}[t]{0.02\textwidth}
  \vspace{-6cm}
  (b)
\end{subfigure}
\adjustbox{minipage=1.3em,valign=t}{\label{sfig:testa}}%
  \begin{subfigure}[t]{\dimexpr.485\linewidth-1.3em\relax}
  \centering
      \hspace*{.8cm}\includegraphics[trim = 1150 120 1000 210, scale=.22]{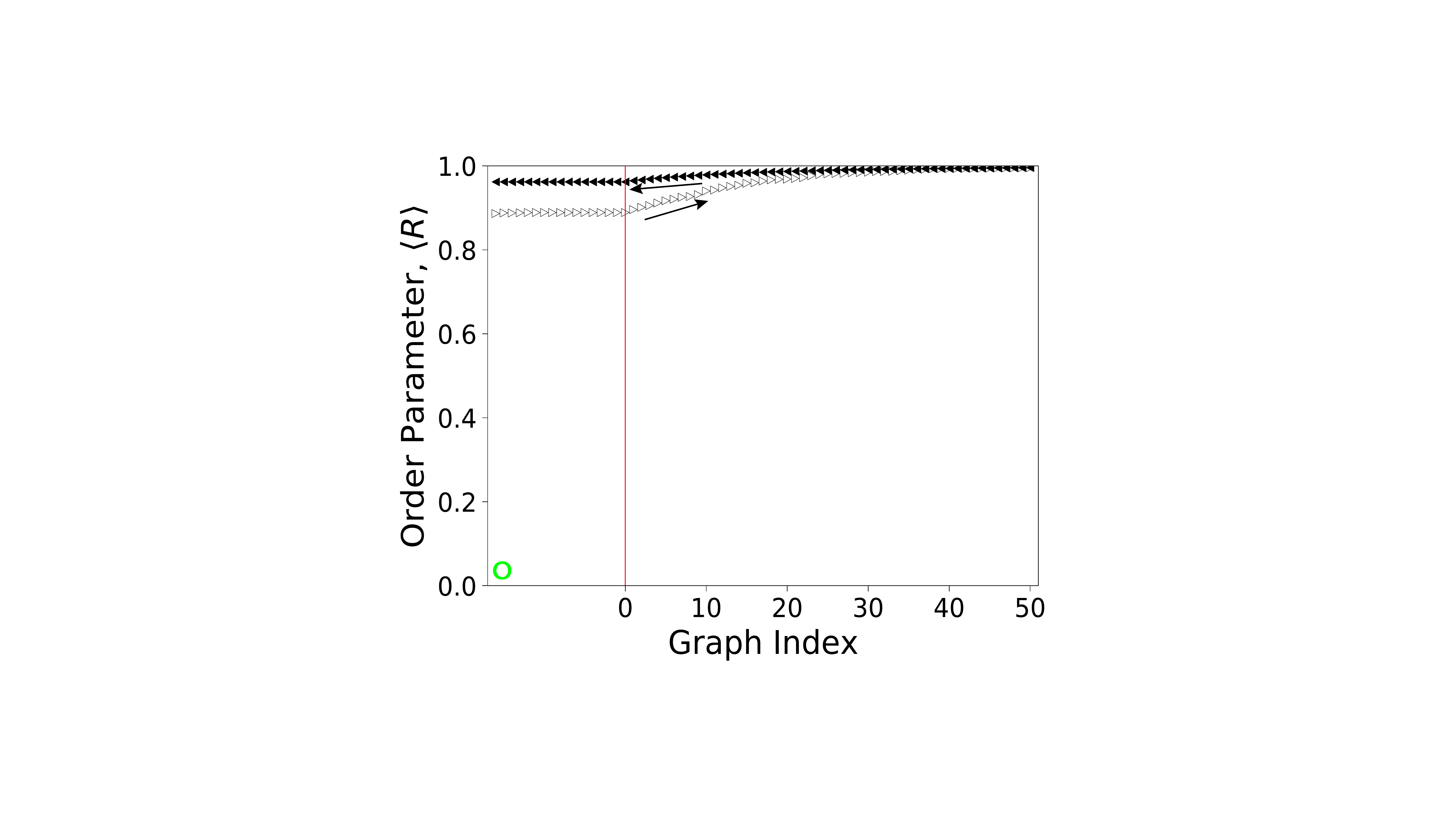}
  \end{subfigure}%
  \\
  \begin{subfigure}[t]{0.02\textwidth}
  \vspace{-6cm}
  (c)
\end{subfigure}
  \adjustbox{minipage=1.3em,valign=t}{\label{sfig:testb}}%
  \begin{subfigure}[t]{\dimexpr.485\linewidth-1.3em\relax}
  \centering
  \hspace*{.8cm} \includegraphics[trim = 1350 -50 1100 0, scale=.15]{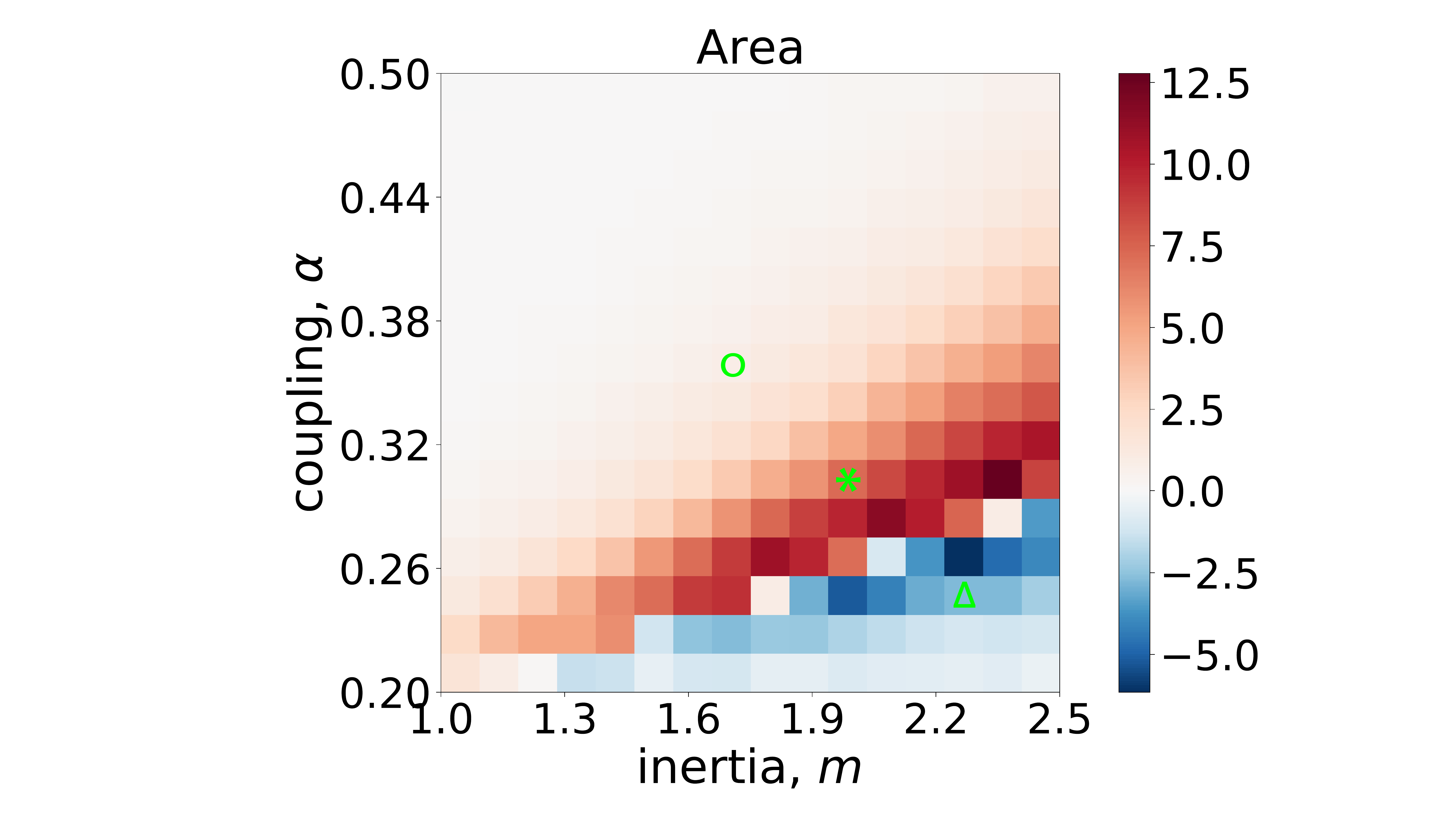}
  \end{subfigure}
  \begin{subfigure}[t]{0.02\textwidth}
  \vspace{-6cm}
  (d)
\end{subfigure}
  \adjustbox{minipage=1.3em,valign=t}{\label{sfig:testa}}%
  \begin{subfigure}[t]{\dimexpr.485\linewidth-1.3em\relax}
  \centering
    \hspace*{.8cm}\includegraphics[trim = 1150 150 1000 210, scale=.22]{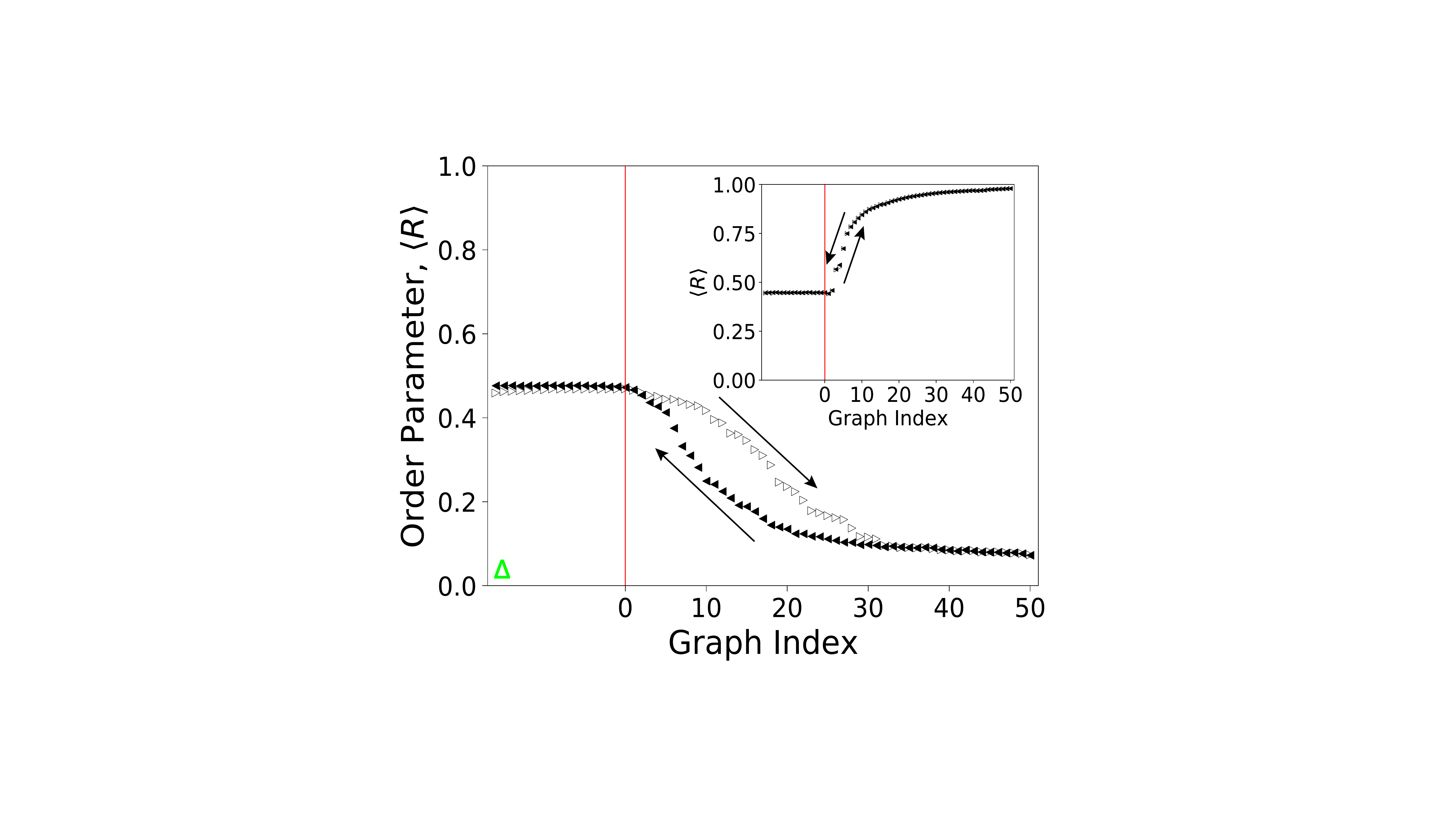}
  \end{subfigure}%

      \caption{\textbf{Effects of fixed-density rewiring from Erdős–Rényi to synchrony-aligned networks and back, as a function of the coupling strength and inertia.} We show the behavior of the order parameter $\langle R \rangle$ as rewiring occurs from Erdős–Rényi $\rightarrow$ synchrony-aligned $\rightarrow$ Erdős–Rényi networks while maintaining a constant density of $\langle k \rangle  = 20$. \emph{(a)} Gain in the level of global synchrony as a result of the rewiring process from Erdős–Rényi $\rightarrow$ synchrony-aligned $\rightarrow$ Erdős–Rényi over a portion of the $\alpha$-$m$ parameter space. \emph{(b)} The time-averaged order parameter $\langle R \rangle$ throughout the rewiring process with parameters $m = 1.7$, $\alpha = 0.36$. \emph{(c)} Area between the forward (Erdős–Rényi $\rightarrow$ synchrony-aligned) and backward (synchrony-aligned $\rightarrow$ Erdős–Rényi) order parameter $\langle R \rangle$ curves over a portion of the $\alpha$-$m$ parameter space. Area is defined such that it is positive when the backward order parameter curve is above the forward curve. Note that we discarded contributions from initial and final transients. \emph{(d)} The time-averaged order parameter $\langle R \rangle$ throughout the rewiring process with parameters $m = 2.3$, $\alpha = 0.24$. The inset reports results from the same process but with the standard Kuramoto model, where the coupling strength has been chosen such that initial levels of synchrony are comparable to the inertial case ($\alpha = 0.169$). Parameter combinations corresponding to Figs.~\ref{fig:5}b, ~\ref{fig:6}b, and ~\ref{fig:6}d are marked accordingly on Figs.~\ref{fig:6}a and ~\ref{fig:6}c. All curves and measures depict averages over 25 instantiations.}
  \label{fig:6}
\end{figure*}

We next quantify how the steady-state synchronization gap $R_{f}^{ss} - R_0^{ss}$ changes over a swath of the inertia-coupling parameter space ($0.2 \leq \alpha \leq 0.5$, $1.0 \leq m \leq 2.5$). Here, $R_{0}^{ss}$ and $R_{f}^{ss}$ represent the initial and final time-averaged order parameters, respectively, after discarding a long transient period (Fig.~\ref{fig:6}a). At high coupling and low inertia, there is little steady-state separation between the initial steady-state order parameter $R_{0}^{ss}$ and the final steady-state order-parameter $R_f^{ss}$. This behavior is expected because oscillators with high coupling and low inertia reach close-to-perfect synchrony on the initial Erdős–Rényi connection topology (Fig.~\ref{fig:12}a); rewiring towards synchrony-aligned networks can therefore only induce a small enhancement of the order parameter (Fig.~\ref{fig:6}b). As detailed further in the following paragraph, low coupling and high inertia also result in negligible steady-state separations $R_f^{ss} - R_{0}^{ss}$ (e.g., the parameter combination denoted by the green triangle in Fig.~\ref{fig:6}d). In contrast, in the regime of moderate coupling and moderate inertia, the network evolution process has a clear sustained effect on the system's collective dynamics as reflected in the steady-state synchronization gap $R_{f}^{ss} - R_0^{ss}$.

To dig deeper into the behavior of the system, we next consider the fact that for some parameter combinations, the Erdős–Rényi $\rightarrow$ synchrony-aligned $\rightarrow$ Erdős–Rényi network evolution could induce a hysteresis loop but not a steady-state synchrony gap. To assess this more nuanced behavior, we calculated the area between the forward and backward order parameter $\langle R \rangle$ curves resulting from Erdős–Rényi $\rightarrow$ synchrony-aligned $\rightarrow$ Erdős–Rényi network evolution, over the same inertia-coupling parameter space  (Fig.~\ref{fig:6}c). For this analysis, the area is defined such that it is positive when the backward order parameter curve is above the forward curve (and we again ignore contributions from initial and final transient periods). Interestingly, we observe a regime at low coupling and high inertia where no steady-state synchrony gap is produced, but hysteresis loops of negative area are formed (Fig.~\ref{fig:6}d). That is, the order parameter actually \textit{decreases} upon rewiring towards synchrony-aligned networks, and then increases back to its initial value along the reverse network evolution pathway.

This type of dynamical trajectory could be a natural consequence of the fact that the derivation of the synchrony alignment function used to produce synchrony-aligned networks employs the approximation of the strong synchrony regime \cite{skardal_2014}. This fact in turn suggests that synchrony-aligned networks could be ineffective in promoting synchronization when synchronizability is already low as a result of parameter choices. However, it is also possible that the ineffectiveness of synchrony-aligned networks (and the emergence of hysteresis loops characterized by negative area) in some parameter regimes is a consequence of inertia rather than initial synchrony levels alone. Indeed, oscillators coupled through synchrony-aligned networks exhibit strong sensitivity to initial synchrony levels when inertia is present (see Supplementary Figs.~\ref{fig:13} and~\ref{fig:14}). To probe this possibility further, we assessed the Erdős–Rényi $\rightarrow$ synchrony-aligned $\rightarrow$ Erdős–Rényi rewiring process using the standard Kuaramoto model, with a coupling strength chosen to make initial synchrony levels comparable to that of the main panel in Fig.~\ref{fig:6}d. Consistent with the idea that inertia is responsible for the ineffectiveness of synchrony-aligned networks in some parameter regimes, we find that standard Kuramoto oscillators with similar levels of initial synchrony still synchronize well when they are rewired towards a synchrony-aligned topology (Fig.~\ref{fig:6}d inset).  

\subsection{Further Network Perturbation}
\begin{figure*}
\begin{subfigure}[t]{0.03\textwidth}
  \vspace{-6cm}
  (a)
\end{subfigure}
  \adjustbox{minipage=1.3em,valign=t}{\label{sfig:testa}}%
  \begin{subfigure}[t]{\dimexpr.485\linewidth-1.3em\relax}
  \centering
     \includegraphics[trim = 1320 180 1000 350, scale=.23]{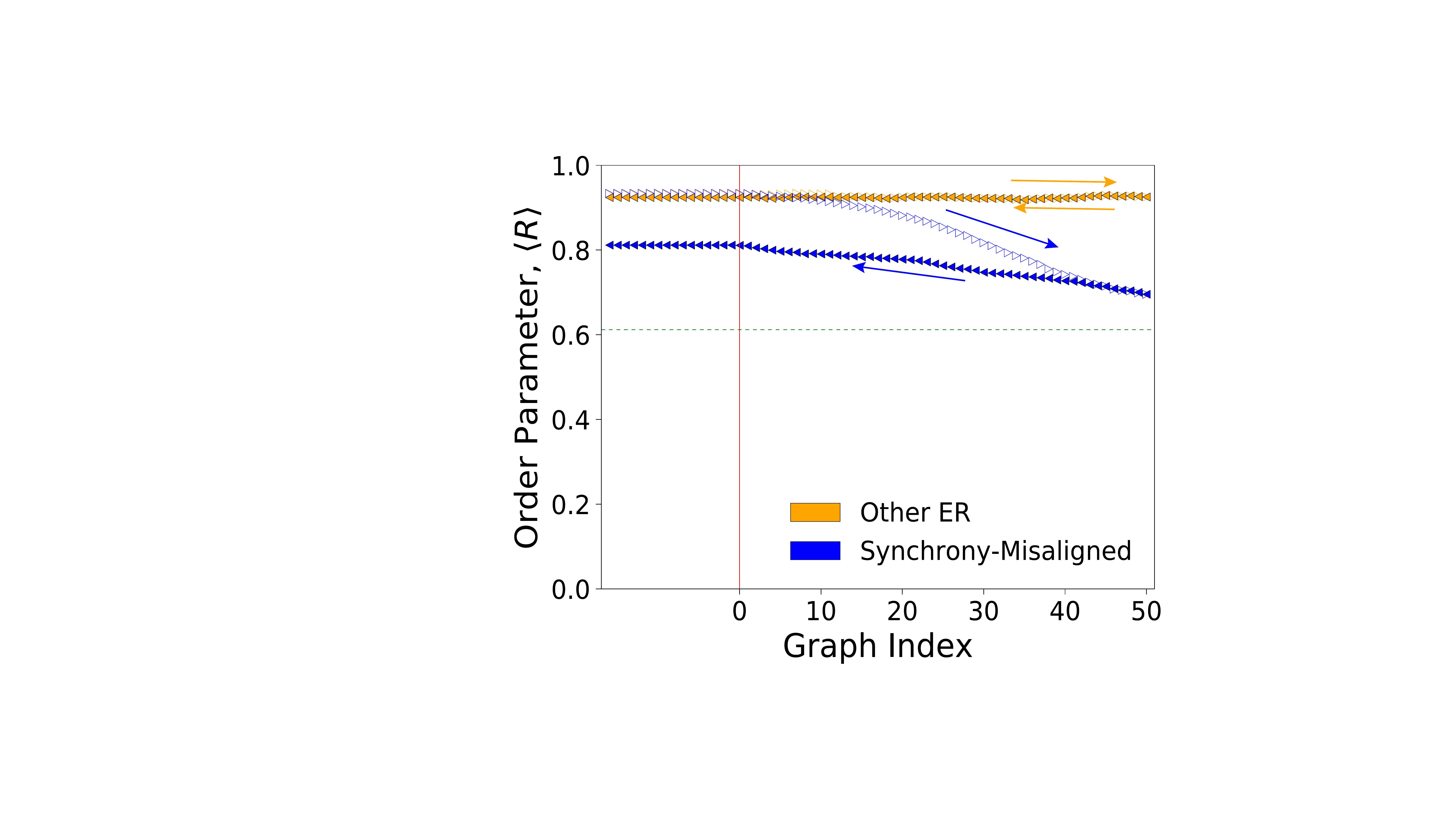}
  \end{subfigure}%
  \adjustbox{minipage=1.3em,valign=t}{\label{sfig:testb}}%
  \begin{subfigure}[t]{0.03\textwidth}
  \vspace{-6cm}
  (b)
\end{subfigure}
  \begin{subfigure}[t]{\dimexpr.485\linewidth-1.3em\relax}
  \centering
    \includegraphics[trim = 1220 220 1000 200, scale=.23]{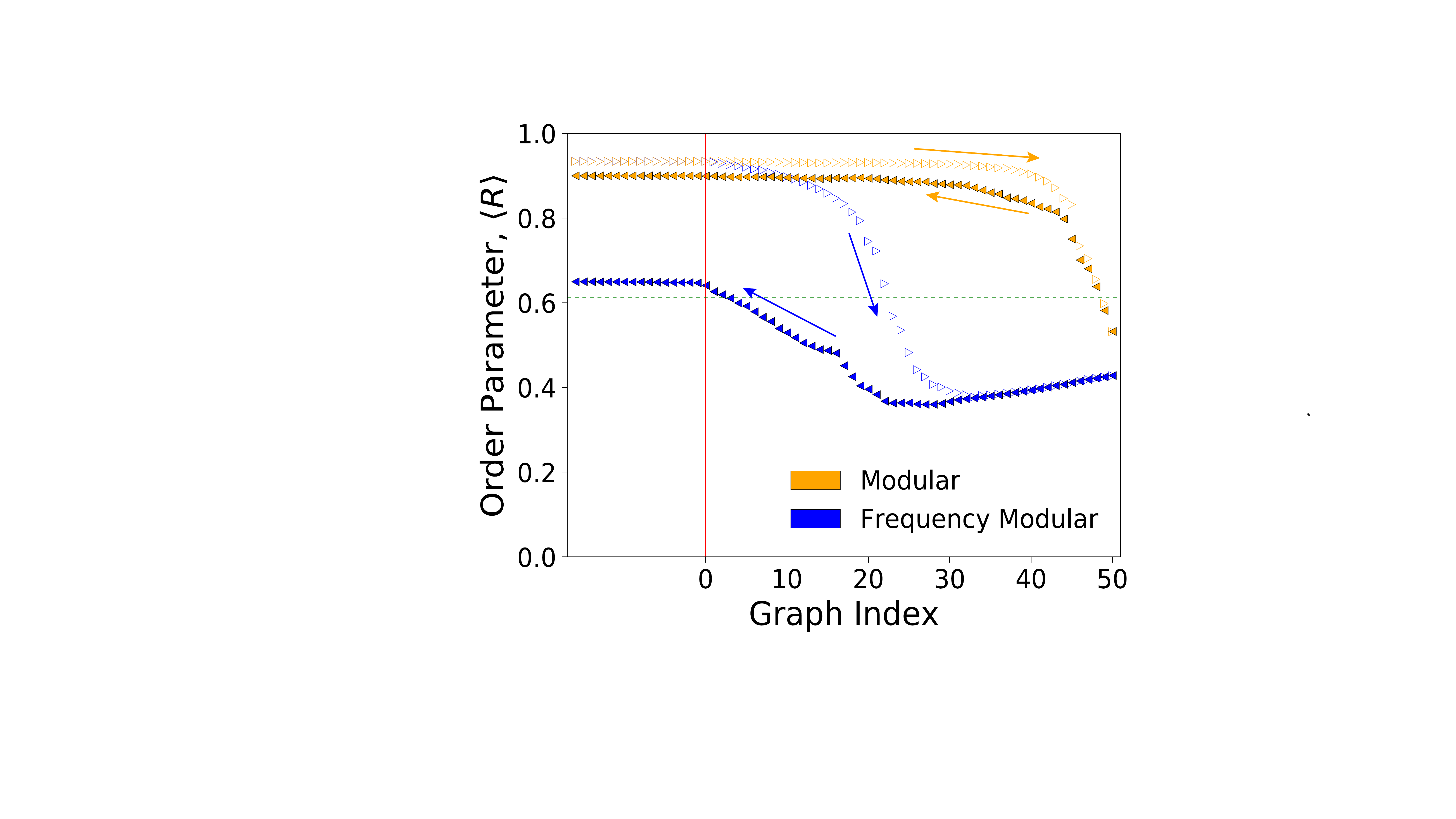}
  \end{subfigure}
  \caption{\textbf{The robustness of gains in global synchrony from network evolution through synchrony-aligned graphs.} We considered the final states of oscillators from Erdős–Rényi $\rightarrow$ synchrony-aligned $\rightarrow$ Erdős–Rényi shown in Fig.~\ref{fig:5}b. We then used these final states as initial conditions for a simulation that began at the same Erdős–Rényi $G_0$ connectivity and evolved toward and away from one of four final network topologies $G_f$. \emph{(a)} Results of simulations in which $G_f$ was set to be another Erdős–Rényi graph (orange), or in which $G_f$ was set to be a synchrony-misaligned graph (blue). \emph{(b)} Results of simulations in which $G_f$ was set to be a modular graph (orange), or in which $G_f$ was set to be a frequency modular graph (blue). Parameters $m = 2$, $\alpha = 0.3$ were used throughout all simulations for both panels. The horizontal dashed-line indicates the baseline level of synchrony obtained with $G_0$ connectivity and random initial conditions; that is, the order parameter prior to any network rewiring. All curves depict averages over 25 instantiations.}
  \label{fig:7}
  \vspace{-.5em}
\end{figure*}

We next sought to quantify the robustness of increases in synchronization due to network rewiring. We began by taking the final, high-synchrony states of inertial oscillators obtained after rewiring towards and away from synchrony-aligned graphs at intermediate coupling (Fig.~\ref{fig:5}b), and using these final states as initial conditions for a set of new simulations. The initial topologies $G_0$ of these new simulations were the original Erdős–Rényi graphs used and the parameters remained fixed at $m =2$, $\alpha = 0.3$. While holding network density constant, we then rewired oscillator connectivity towards and away from one of four final network topologies $G_f$: 1) other Erdős–Rényi graphs, 2) synchrony-misaligned graphs, 3) random modular graphs, or 4) frequency modular graphs (see Supplementary Materials for details on graph construction). These final network structures were chosen so as to assess the level of topological perturbation needed to effectively desynchronize systems of inertial oscillators in a high-synchrony state induced by a particular network evolution history. 

We hypothesized that further network evolution toward and away from other Erdős–Rényi graphs would have little effect on global synchrony, and would preserve most of the prior synchrony gains. In contrast, we expected that networks with modular organization may effectively erase global synchrony gains resulting from a specific path of network evolution. In particular, we conjectured that frequency modular graphs---graphs created by assigning oscillators with similar natural frequencies to the same module---would be most effective in perturbing rewiring-induced gains in global synchrony. Consistent with intuition, we found that rewiring towards other Erdős–Rényi graphs had little-to-no effect on levels of synchrony (Fig.~\ref{fig:7}a). In every numerical experiment, the system remained at the enhanced synchrony level acquired under Erdős–Rényi $\rightarrow$ synchrony-aligned $\rightarrow$ Erdős–Rényi network evolution, with little deviation throughout both the forward and backward rewiring trajectories. This finding suggests that gains in synchrony due to network rewiring through synchrony-aligned graphs are quite robust to further random network perturbations. 

Still, it remains unclear as to which topologies might be able to desynchronize inertial oscillators with synchrony gains resulting from a particular network evolution history. To probe this question further, one natural idea is to use synchrony-\textit{misaligned} graphs for the $G_f$ network structures (blue curves in Fig.~\ref{fig:5}a). Such networks are constructed by maximizing (rather than minimizing) the synchrony alignment function (Eq. 4), and thus should theoretically be quite difficult to synchronize. We found that rewiring trajectories towards the synchrony-misaligned graphs induced partial desynchronization of the oscillators. Interestingly, though, we observed clear irreversibility in the order parameter as we rewired from the synchrony-misaligned graphs back to the original Erdős–Rényi graphs. Specifically, the system did not fully return to the baseline synchrony level obtained with random initial conditions (green dotted line in Fig.~\ref{fig:7}a).

For our final analysis, we wished to investigate whether networks that promote local synchrony can effectively reset gains in global synchrony resulting from network history. To do so, we considered modular networks, which have topologies known to favor local synchrony over global synchrony. Rewiring towards both random modular and frequency modular $G_{f}$ graphs greatly reduced the global synchrony of the oscillators (see forward trajectories in Fig.~\ref{fig:7}b). However, only evolution towards frequency modular graphs gave rise to effects that remained even after Erdős–Rényi $G_0$ connectivity was recovered (see backward trajectories in Fig.~\ref{fig:7}b). This behavior might occur because, in addition to discouraging global synchronization, frequency modular graphs are more prone to allowing oscillators to evolve onto the cluster synchronization manifold, resetting much of the history of global synchrony (see Supplementary Fig.~\ref{fig:15}). Note also that rewiring towards frequency-modular graphs yields a slightly lower level of global synchrony than rewiring to modular graphs, which may also play a role in determining the final level of synchrony after rewiring back to the Erdős–Rényi networks.

In sum, our results indicate that the extent to which enhanced synchrony is maintained after further network rewiring depends on more than just how effectively the $G_f$ topology reduces global synchrony during its presence. In particular, global synchrony while $G_f$ topology was present was higher for synchrony-misaligned graphs than for random modular graphs, but as the system returned to its initial topology along the backward transition, the synchrony-misaligned pathway ultimately led to more sustained desynchronization. This pattern of findings suggests that the specific pathway of network evolution can play a key role in modulating the collective dynamics of coupled oscillators, beyond just the immediate effects that different network structures have on synchrony.   

\section{Discussion}
\label{sec:IV}

In this paper, we investigated how various routes of network evolution affect systems of coupled oscillators. Networks of standard Kuramoto oscillators are monostable \cite{Esmaeili_2017, Labavic_Meyer-Ortmanns_2017}. Therefore, under generic conditions they are unaffected by network rewiring processes; the post-transient global synchrony of oscillators at a given time are a function of just the network structure present at that time. However, it may not be the case that this path-independent behavior persists in systems of inherently multistable oscillators. To probe the question of whether multistability in the dynamics of individual oscillators gives rise to path-dependent behavior under network rewiring, we used the inertial Kuramoto model, which adds an inertial term to the standard Kuramoto model and consequently exhibits sensitivity to initial conditions \cite{olmi_2015, Jaros_Brezetsky_Levchenko_Dudkowski_Kapitaniak_Maistrenko_2018}. Prior work has shown that systems of inertial oscillators can exhibit hysteretic synchronization transitions as the coupling strength is increased and then decreased \cite{olmi_navas_boccaletti_torcini_2014}. For networked oscillators, this tuning of the coupling strength can be regarded as a global scaling of the strength of connections that leaves the network topology intact. 

However, in many systems, it is the network organization itself---i.e., where edges exist or do not exist---that is dynamic, rather than the overall strength of each connection \cite{Laurent_Saramaki_Karsai_2015, Calhoun_Miller_Pearlson_Adali_2014, Lebre_Becq_Devaux_Stumpf_Lelandais_2010}. In these cases, it then becomes interesting to ask whether networks of inertial oscillators exhibit path-dependent dynamics induced by changes in network organization alone. To answer this question, we developed a network rewiring procedure that isolates the effects of network evolution history. Specifically, we evolved the network connectivity of systems of coupled inertial (and non-inertial) oscillators towards a pre-specified final network structure, and then we reversed the rewiring process along the same path. In this way, any path-dependent synchronization behavior would be reflected as asymmetries of the order parameter between the forward and backward network rewiring trajectories. 

We first investigated the effects of slowly increasing the network density of random graph topology and then reversing the evolution until the original graph was recovered. For oscillators with moderate inertia and coupling, we found that this density-varying process could induce hysteretic synchronization behavior, with oscillators preferring to stay in a more globally synchronized state for more of the backward rewiring process than the forward rewiring process. This finding is in line with the aforementioned work on hysteretic behavior of inertial Kuramoto oscillators with tuning of the coupling strength \cite{olmi_navas_boccaletti_torcini_2014}; it is natural to expect that increasing the density of connections in non-sparse random networks will yield similar effects to that of globally increasing the strength of connections between oscillators. Expanding upon this understanding, the formation of a hysteresis loop as network density is varied demonstrates the existence of situations where identical network connectivity patterns give rise to different oscillator dynamics due to differences in network evolution history alone. Going a step further, we then analyzed the case of varying oscillator network density at high coupling, where we uncovered a qualitatively unique form of path-dependent behavior in which network rewiring results in irreversible gains in global synchrony.

To isolate the role of network topology from density-driven effects in path-dependent behavior, we next studied how inertial oscillators behave when their network topology is rewired at constant density. Specifically, we generated networks known to be highly synchronizable for the standard Kuramoto model, and analyzed the effects of constant-density rewiring of initially randomly-coupled networks of inertial oscillators toward and then away from these synchrony-aligned networks. Notably, we found that even when density is held constant throughout the network rewiring process, the dynamics of inertial oscillators can depend on previous network evolution history, and that this path-dependence is significant for a considerable portion of the inertia-coupling parameter space. In addition, gains in synchrony due to constant-density rewiring were robust to a number of subsequent network perturbations, with near-complete reversal of effects occurring only in the extreme case of further rewiring towards highly cluster-synchronizable networks. Collectively, these results demonstrate that variations in topology alone can drive path-dependent dynamics of inertial oscillators, and that the resulting effects typically persist even with further network evolution. 

\noindent \textbf{Opportunities for extensions and expansions.} Our results prompt a number of interesting directions for further investigation, particularly related to the nature of the graphs studied, the nature of the rewiring process, and expansions to other formalisms and models. First we note that we considered networks with binary connectivity only. While a reasonable place to begin, it would be interesting in the future to study inertial oscillators evolving on time-varying \emph{weighted} networks as well. In particular, the weights could be made to vary continuously in time \cite{cumin_unsworth_2007, Leander_Lenhart_Protopopescu_2015, Petkoski_Stefanovska_2012}, possibly making stability analysis of oscillator dynamics evolving along different network evolution pathways more analytically tractable. Such an extension could increase the relevance of our observations to real world systems, which are typically characterized by a high density of edges whose weights can vary over several orders of magnitude \cite{Serrano_Boguna_Vespignani_2009, Farahani_Karwowski_Lighthall_2019, Jin_Girvan_Newman_2001, Saxena_Iyengar_2016}.

Second, we studied the effects of rewiring an initial network topology towards a final network topology, where the necessary edges to be moved were rewired in a random order. A future study could consider whether the order in which edges are rewired plays a significant role in the development of path-dependent behavior. Prior work has shown that enforcing certain relationships between pairwise differences in the natural frequencies of oscillators and their coupling patterns may promote more complex oscillator dynamics, such as explosive synchronization \cite{Leyva_Navas_Sendina-Nadal_Almendral_Buldu_Zanin_Papo_Boccaletti_2013, Leyva_Sendina-Nadal_Almendral_Navas_Olmi_Boccaletti_2013}. Therefore, it is possible that first placing edges between oscillators with the least similar natural frequencies may greatly affect how synchronization develops in both the forward and backward rewiring processes. Moreover, while the rewiring process we used was convenient for illustrating potential path-dependent behavior, it was controlled in the sense that only the edges that ultimately needed to be added or removed to arrive at some final network structure were rewired, and each relevant edge was only altered once. It would be interesting to see how inertial oscillators behave when the rewiring process occurs in a more organic manner, such as by allowing for all edges to be added and pruned repeatedly while still maintaining that some pre-determined final network structure is eventually reached. 

A third possible area for future study is to consider how path-dependence arises in a system of inertial oscillators adhering to an adaptive rewiring scheme, where the states of the oscillators themselves inform the network rewiring process \cite{papadopoulos_2017, Aoki_Aoyagi_2009, Aoki_Aoyagi_2011,Petkoski_Stefanovska_2012, Gross_Blasius_2008, Gross_Sayama_2009, Zhu_Zhao_Yu_Zhou_Wang_2010}. In particular, investigating path-dependence in systems of inertial oscillators under Hebbian or anti-Hebbian adaptive rewiring \cite{Niyogi_English_2009, Bronski_He_Li_Liu_Sponseller_Wolbert_2017, Skardal_Taylor_Restrepo_2014} may be insightful for understanding whether network evolution path-dependence plays a significant role in the development of neuronal networks. It is well known that the human brain undergoes a variety of structural changes during development \cite{tang2017developmental,cornblath2019sex,baum2020development}, not only in synaptic density but also in topological characteristics such as degree heterogeneity, clustering, and modularity \cite{Khundrakpam_Reid_Brauer_Carbonell_Lewis_Ameis_Karama_Lee_Chen_Das_et_al_2013b, Gao_Alcauter_Smith_Gilmore_Lin_2015}. Such changes complicate any inferences drawn from the existing levels of synchronization, which may be both a function of the current network topology and a function of the network's developmental history.

Another interesting avenue for future work would be to examine how multilayer oscillator networks behave in the presence of inertia and network rewiring. Prior work has shown that multilayer oscillator networks exhibit a variety of rich behaviors \cite{bianconi2018multilayer}. For example, explosive synchronization may occur in two-layer networks with adaptive coupling \cite{Zhang_Boccaletti_Guan_Liu_2015,dsouza2019explosive}. Such behavior has even been shown to occur when an oscillator network is coupled to an entirely different dynamical process, such as a nutrient transport layer \cite{Nicosia_Skardal_Arenas_Latora_2017}. Future work could fruitfully investigate how multilayer oscillator networks evolve in the presence of inertia, and examine the role of path-dependence in multilayer oscillator networks with dynamic connectivity.

Finally, it is worth noting that other variants of the Kuramoto model can also exhibit multistability, such as the Kuramoto-Sakaguchi model with time-delayed coupling \cite{pazo_montbrio_2009, metivier_gupta_2019, coutinho_goltsev_dorogovtsev_mendes_2013, Yeung_Strogatz_1999}. It would thus also be insightful to investigate the interplay between multistability arising from these alternative means and the network evolution of oscillator connectivity.\\  

\noindent \textbf{Conclusion.} 
Discerning the effects of dynamic network organization on the collective behavior of coupled dynamical subunits remains an important area of study, with implications for a number of physical and biological systems \cite{motter_myers_anghel_nishikawa_2013, buck_1988, cumin_unsworth_2007, Laurent_Saramaki_Karsai_2015, Calhoun_Miller_Pearlson_Adali_2014, Lebre_Becq_Devaux_Stumpf_Lelandais_2010}. To understand whether oscillators coupled through time-varying networks can be significantly affected by the history of the coupling network, or if only the final network structures obtained from the network evolution process are relevant, we have studied the impact of various network rewiring pathways on systems of inertial Kuramoto oscillators. We have observed many situations throughout this study where markedly different oscillator dynamics and levels of synchrony can arise from the same network structure solely due to differences in prior connectivity patterns. In particular, we have found that network rewiring can drive hysteretic synchronization behavior of inertial oscillators via increasing and decreasing network density, and that even changes in topology alone from constant-density network rewiring can induce path-dependent behavior. These findings demonstrate that beyond the overdamped limit, pathways of network evolution themselves play an important role in regulating the behavior of networks of coupled subunits, and require consideration when studying the dynamics of systems evolving over complex networks, and when devising strategies for their control. 

\section{Acknowledgments}
WQ acknowledges support from the Vagelos program at the University of Pennsylvania. FP and DSB acknowledge support from the National Science Foundation, through a collaborative grant funding mechanism (IIS-1926757). LP, ZL, KW, and DSB also acknowledge further support from the Paul G. Allen Family Foundation, the National Science Foundation (PHY15-54488, DMR-1420530), and the Army Research Office (W911NF-16-1-0474, W011MF-191-244). The content is solely the responsibility of the authors and does not necessarily represent the official views of any of the funding agencies.

\section{Citation Diversity Statement} 
Recent work in several fields of science has identified a bias in citation practices such that papers from women and other minorities are under-cited relative to the number of such papers in the field \cite{Dworkin2020, maliniak2013gender, caplar2017quantitative, chakravartty2018communicationsowhite, YannikThiemKrisF.SealeyAmyE.FerrerAdrielM.Trott2018, dion2018gendered}. Here we sought to proactively consider choosing references that reflect the diversity of the field in thought, form of contribution, gender, and other factors. We obtained predicted gender of the first and last author of each reference by using databases that store the probability of a name being carried by a woman \cite{Dworkin2020,cleanbib}. By this measure (and excluding self-citations to the first and last authors of our current paper), our references contain 9.4\% woman(first)/woman(last), 7.8\% man/woman, 15.6\% woman/man, 56.2\% man/man, and 10.94\% unknown categorization. This method is limited in that a) names, pronouns, and social media profiles used to construct the databases may not, in every case, be indicative of gender identity and b) it cannot account for intersex, non-binary, or transgender people. We look forward to future work that could help us to better understand how to support equitable practices in science.

\newpage
\section{Supplementary Material}

\noindent \textbf{Generating synchrony-aligned networks}: We generated synchrony-aligned networks of a given average degree $\langle k \rangle$ via a hill-climbing algorithm \cite{skardal_2014} as follows: starting with an Erdős–Rényi graph $G$ with average degree $\langle k \rangle$, an edge is deleted at random and replaced with an edge between two randomly chosen nonadjacent nodes, producing $G'$. If $J(\boldsymbol{\omega}, L') < J(\boldsymbol{\omega}, L)$, then $G'$ is accepted; otherwise, the original graph $G$ is retained. This process was repeated for $2 \times 10^4$ iterations until a synchrony-aligned graph $G^{\star}$ was obtained. Synchrony-misaligned networks were generated following the same procedure, but instead sought to maximize $J(\boldsymbol{\omega}, L)$.

~\\

\noindent \textbf{Generating modular networks}: Random modular graphs were generated by randomly assigning each oscillator to one of five modules, and then preferentially adding edges within modules so that 90\% of edges were intra-modular. Frequency modular graphs were generated similarly, but instead assigned modules to oscillators based on the similarity of their natural frequencies. This assignment was done by dividing the interval $[\omega_{min}, \omega_{max}]$ evenly into five intervals of the same length, where $\omega_{min}$ and $\omega_{max}$ correspond to the minimum and maximum natural frequency, respectively. Each oscillator was then assigned to a module corresponding to the one of five intervals that its natural frequency fell within.

~\\

\noindent \textbf{Assessing cluster synchrony:} We  quantify the time-averaged pairwise synchrony between oscillators $i$ and $j$ in Figs. \ref{fig:15}c and \ref{fig:15}d by 
\begin{equation}
\langle R_{i,j} \rangle = \frac{1}{T}\left|\int_{T_R}^{T_R + T} \frac{1}{2}(e^{i\theta _j(t)} + e^{i\theta_i(t)}) dt\right|.
\end{equation}

These values were computed while the modular or frequency modular $G_f$ topology was present during Erdős–Rényi $\rightarrow G_f \rightarrow$ Erdős–Rényi network evolution. As in Fig.~\ref{fig:7}b, the initial conditions were obtained from the high-synchrony states resulting from Erdős–Rényi $\rightarrow$ synchrony-aligned $\rightarrow$ Erdős–Rényi evolution (Fig.~\ref{fig:5}b).

\hspace{-3cm}\begin{figure*}
\begin{subfigure}[t]{0.03\textwidth}
  \vspace{-9.2cm}
  \hspace{-7cm}
  (a)
\end{subfigure}
  \adjustbox{minipage=1.3em,valign=t}{\label{sfig:testa}}%
  \begin{subfigure}[t]{\dimexpr.485\linewidth-1.3em\relax}
  \centering
     \includegraphics[width = 1.4\columnwidth, trim = 700 200 400 200]{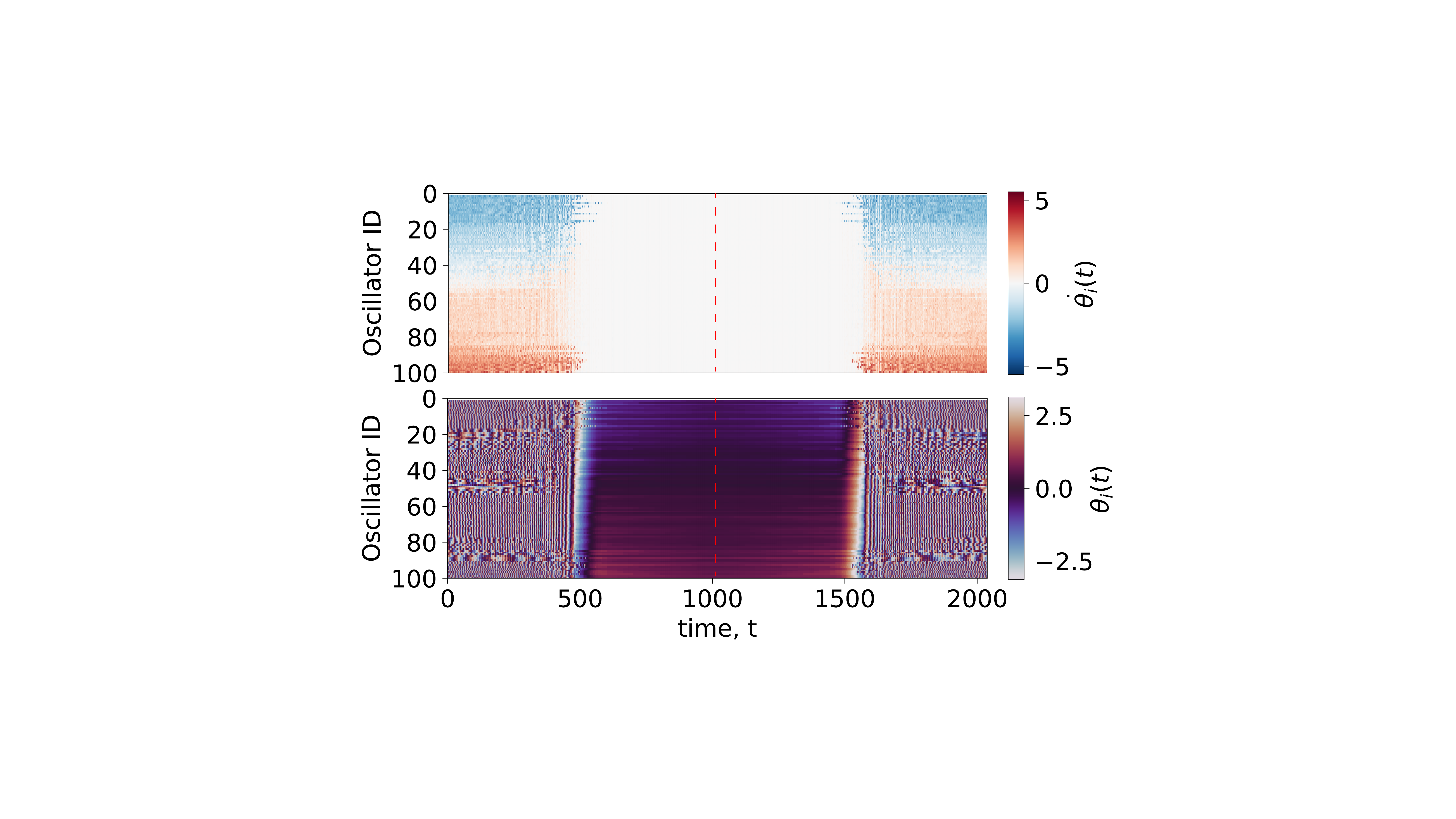}
  \end{subfigure}%
  \\
  \begin{subfigure}[t]{0.03\textwidth}
  \vspace{-9.2cm}
  \hspace{-7cm}
  (b)
\end{subfigure}
  \adjustbox{minipage=1.3em,valign=t}{\label{sfig:testb}}%
  \begin{subfigure}[t]{\dimexpr.485\linewidth-1.3em\relax}
  \centering
    \includegraphics[width = 1.4\columnwidth, trim = 700 200 400 200]{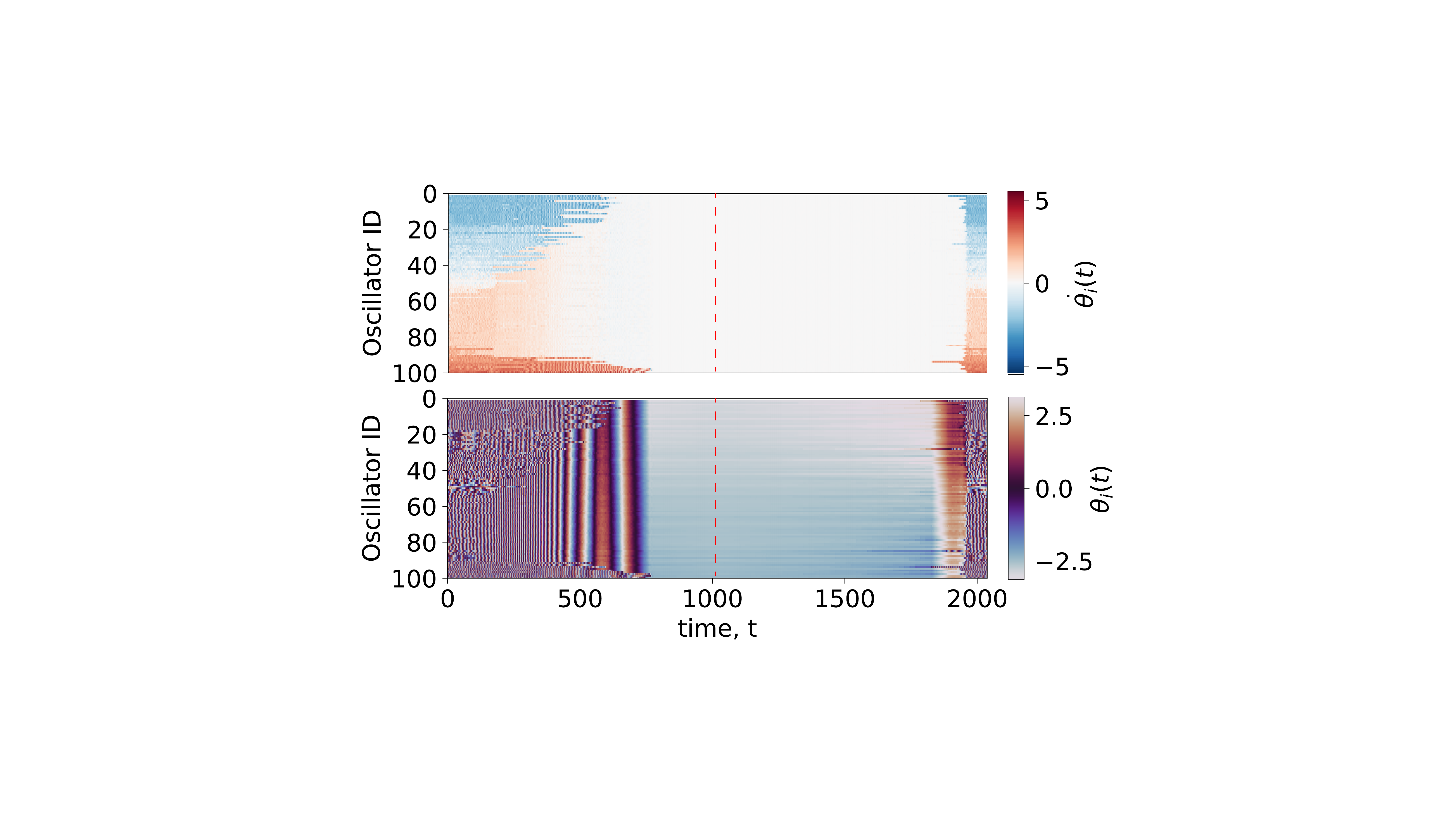}
  \end{subfigure}
  \caption{\textbf{Oscillator dynamics while varying network density.} Single-instance examples of the phases and frequencies of the oscillators vs. time $t$ as network density is first increased ($\langle k \rangle = 10 \rightarrow 40$) and then decreased ($\langle k \rangle = 40 \rightarrow 10$), with and without inertia. The red dotted line indicates the start of reverse network evolution. The oscillators are sorted in ascending order of natural frequencies, with each row corresponding to the time series of a particular oscillator. For visual clarity, these examples were produced on a faster rewiring timescale than described in the main text ($l = 1 \times 10^3$), with every tenth phase and frequency shown. The conditions for each simulation were chosen to make minimum and maximum levels of synchrony comparable: (a) without inertia ($m = 0$, $\alpha = 0.15$) (b) with inertia ($m = 2$, $\alpha = 0.3$).}
  \label{fig:8}
\end{figure*}

\begin{figure*}
\begin{subfigure}{0.03\textwidth}
  \vspace{-19cm}
  \hspace{-7cm}(a)
\end{subfigure}
  \adjustbox{minipage=1.3em,valign=t}{\label{sfig:testa}}%
  \begin{subfigure}[t]{\dimexpr.485\linewidth-1.3em\relax}
  \centering
     \includegraphics[width = 1.5\columnwidth, trim = 700 200 400 200]{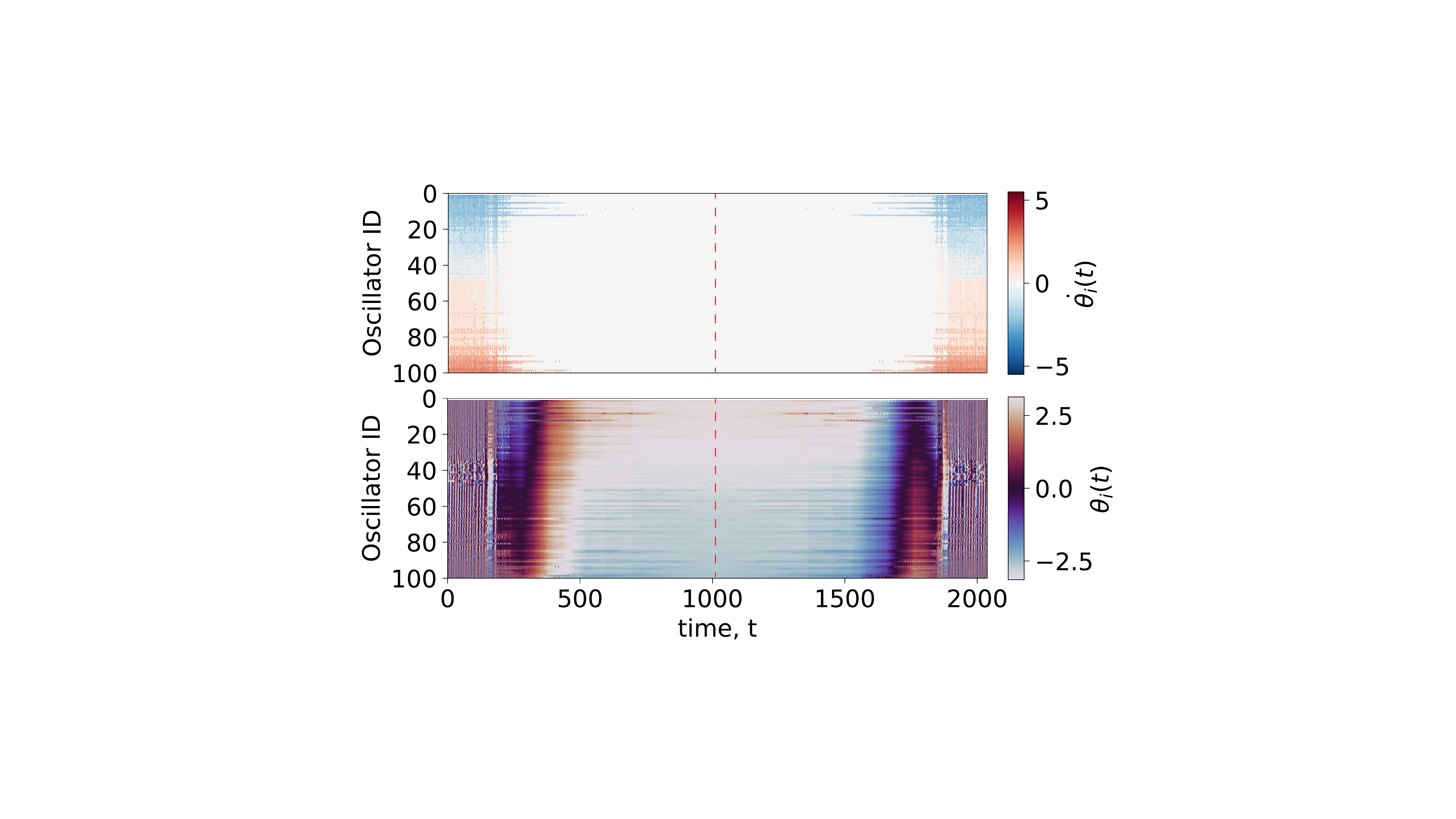}
  \end{subfigure}%
  \\
  \begin{subfigure}[t]{0.03\textwidth}
  \vspace{-9.7cm}
  \hspace{-7cm}
  (b)
\end{subfigure}
  \adjustbox{minipage=1.3em,valign=t}{\label{sfig:testb}}%
  \begin{subfigure}[t]{\dimexpr.485\linewidth-1.3em\relax}
  \centering
    \includegraphics[width = 1.5\columnwidth, trim = 700 200 400 200]{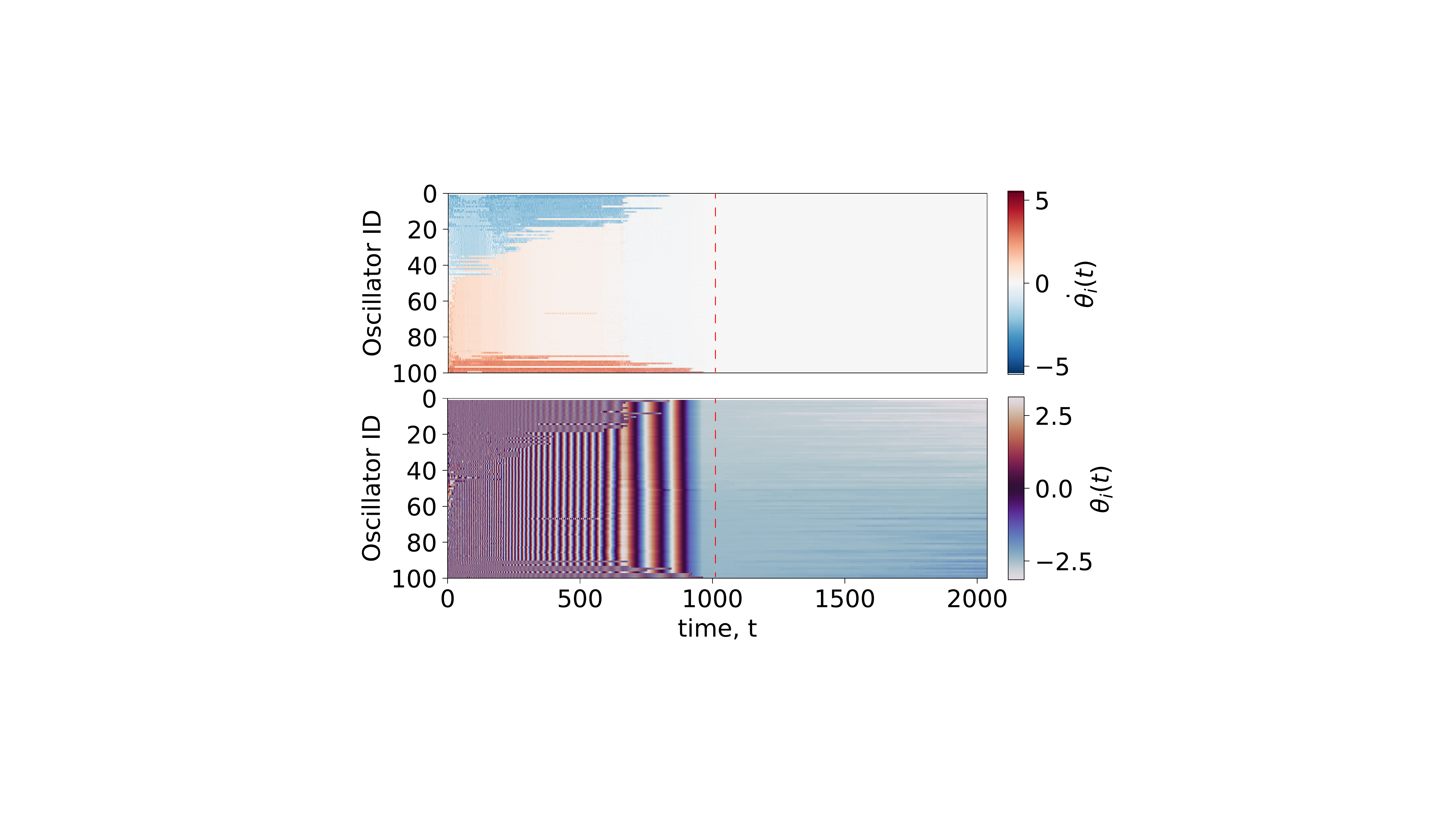}
  \end{subfigure}
  \caption{\textbf{Oscillator dynamics while undergoing constant-density rewiring.} Single-instance examples of the phases and frequencies of the oscillators vs. time $t$ as oscillators undergo Erdős–Rényi $\rightarrow$ synchrony-aligned $\rightarrow$ Erdős–Rényi rewiring at constant density ($\langle k \rangle = 20$), with and without inertia. The red dotted line indicates the beginning of reverse network evolution. The oscillators are sorted in ascending order of natural frequencies, with each row corresponding to the time series of a particular oscillator. For visual clarity, these examples were produced on a faster rewiring timescale than described in the main text ($l = 1 \times 10^3$), with every tenth phase and frequency shown. The conditions for each simulation were chosen to make minimum and maximum levels of synchrony comparable: (a) without inertia ($m = 0$, $\alpha = 0.15$) (b) with inertia ($m = 2$, $\alpha = 0.3$).}
  \label{fig:9}
\end{figure*}

\begin{figure*}
\begin{subfigure}[t]{0.03\textwidth}
  \vspace{-5cm}
  (a)
\end{subfigure}
  \adjustbox{minipage=1.3em,valign=t}{\label{sfig:testa}}%
  \begin{subfigure}[t]{\dimexpr.485\linewidth-1.3em\relax}
  \centering
     \includegraphics[trim = 950 250 1000 400, scale=.24]{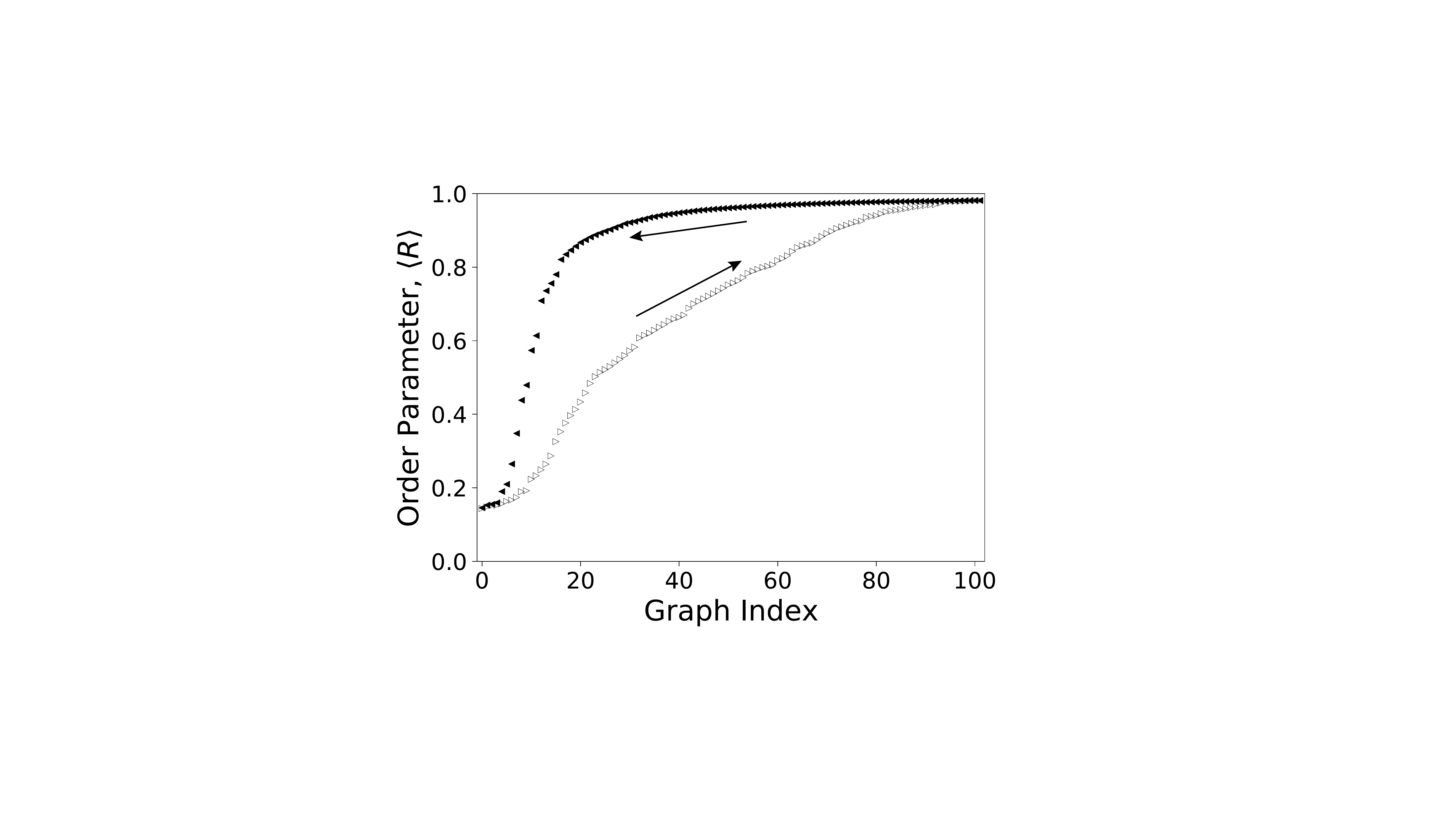}
  \end{subfigure}%
  \begin{subfigure}[t]{0.03\textwidth}
  \vspace{-5cm}
  (b)
\end{subfigure}
  \adjustbox{minipage=1.3em,valign=t}{\label{sfig:testb}}%
  \begin{subfigure}[t]{\dimexpr.485\linewidth-1.3em\relax}
  \centering
    \includegraphics[trim = 950 250 1000 400, scale=.24]{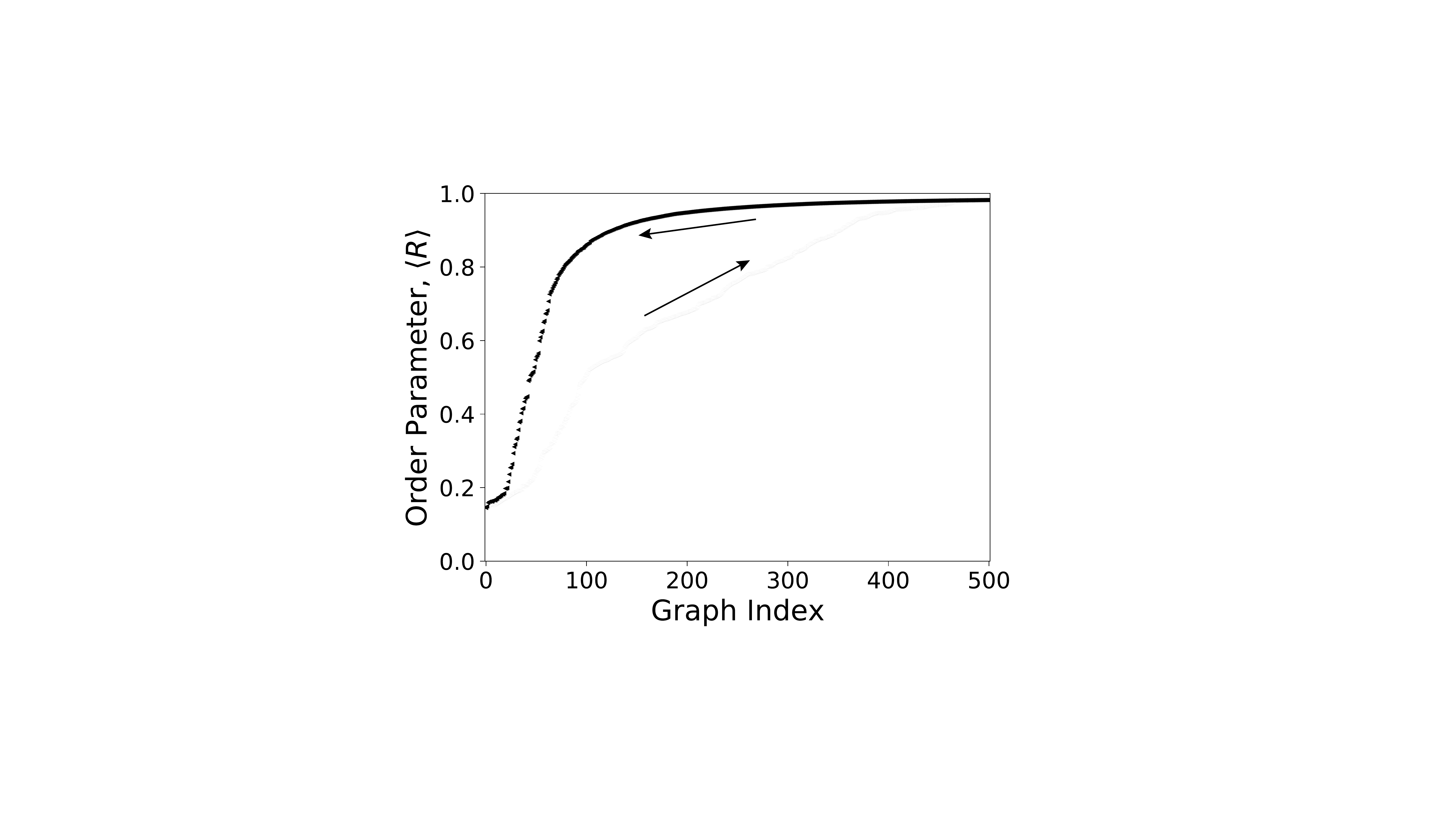}
  \end{subfigure}
  \caption{\textbf{Varying network density at different timescales.} The time-averaged order parameter $\langle R \rangle$ as network density is first increased ($\langle k \rangle = 10 \rightarrow 40$) and then decreased ($\langle k \rangle = 40 \rightarrow 10$) at different values of $l$ and $f$, where $l \times f$ is kept constant to fix the total integration time. Both figures were produced with parameters $m = 2$, $ \alpha = 0.3$ (i.e., with inertia). (a) Here, $l = 2.5 \times 10^4$ and $f = 100$. (b) Here, $l = 5 \times 10^3$ and $f = 500$. All curves depict averages over 25 instantiations.}
  \label{fig:10}
\end{figure*}

\begin{figure*}[b]
\begin{subfigure}[t]{0.03\textwidth}
  \vspace{-5.4cm}
  (a)
\end{subfigure}
  \adjustbox{minipage=1.3em,valign=t}{\label{sfig:testa}}%
  \begin{subfigure}[t]{\dimexpr.485\linewidth-1.3em\relax}
  \centering
     \includegraphics[trim = 1000 200 1000 250, scale=.22]{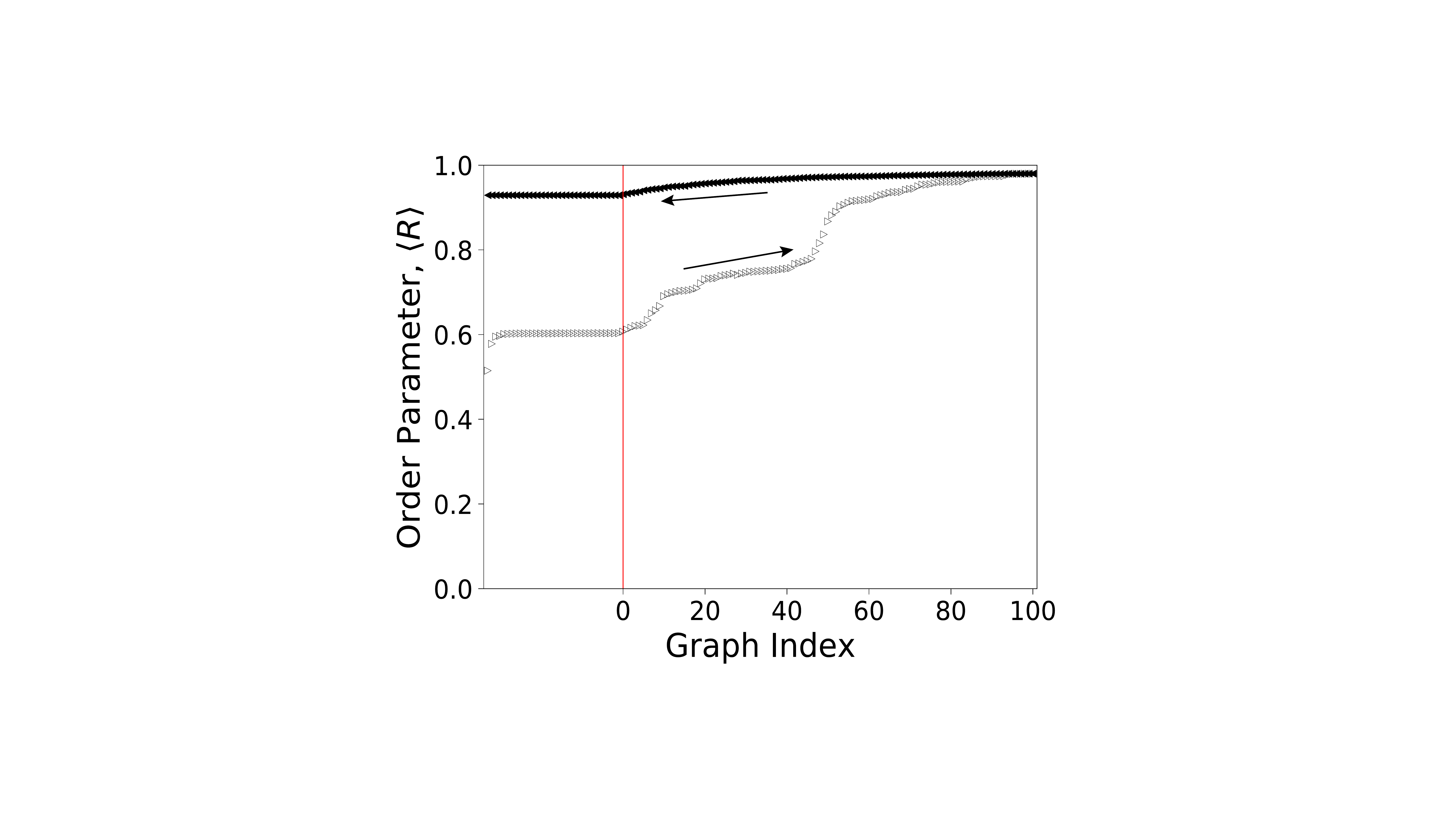}
  \end{subfigure}%
  \begin{subfigure}[t]{0.03\textwidth}
  \vspace{-5.4cm}
  (b)
\end{subfigure}
  \adjustbox{minipage=1.3em,valign=t}{\label{sfig:testb}}%
  \begin{subfigure}[t]{\dimexpr.485\linewidth-1.3em\relax}
  \centering
    \includegraphics[trim = 1020 200 1000 250, scale=.22]{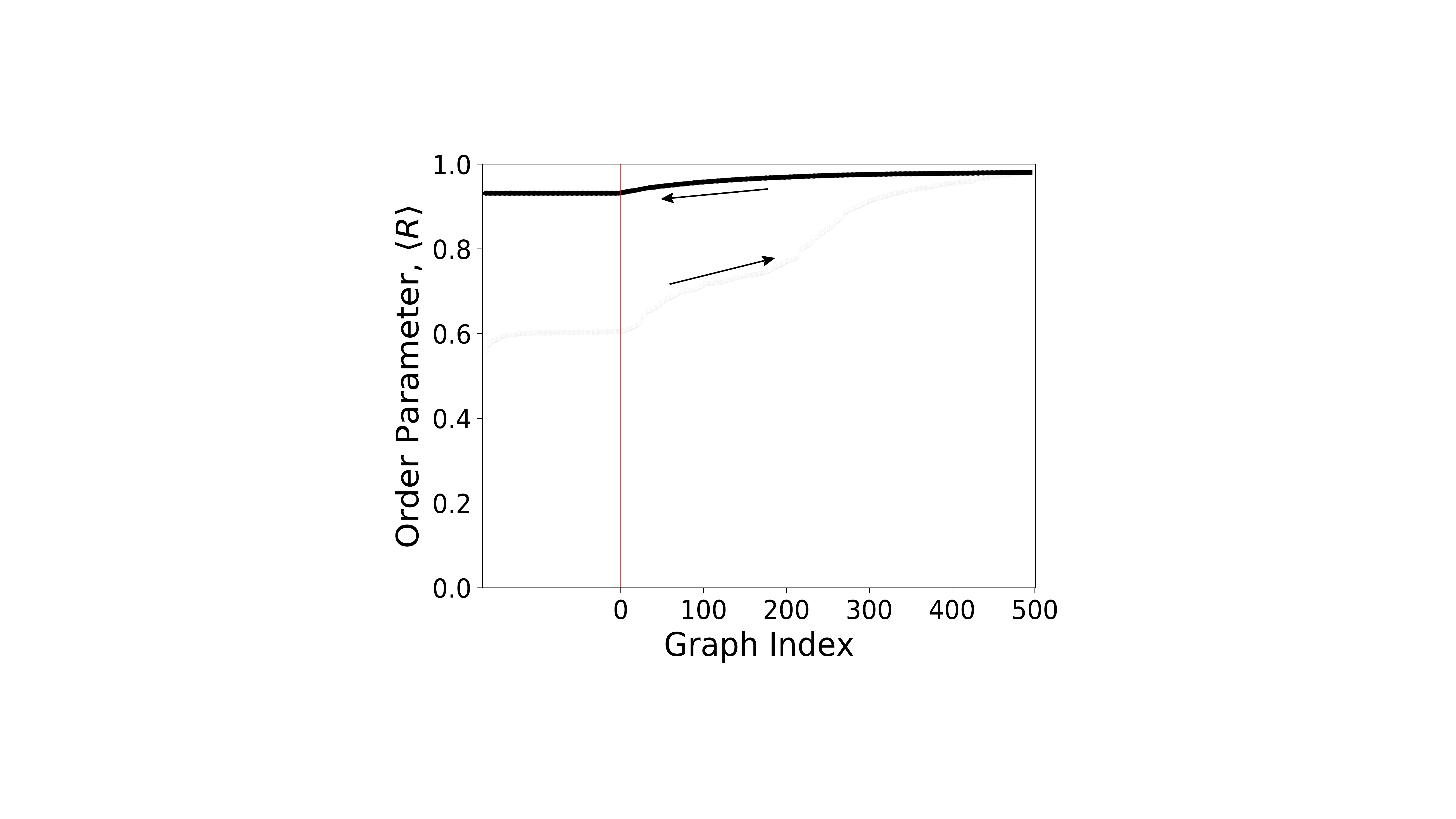}
  \end{subfigure}
  \caption{\textbf{Constant-density rewiring at different timescales.} The time-averaged order parameter $\langle R \rangle$ as Erdős–Rényi $\rightarrow$ synchrony-aligned $\rightarrow$ Erdős–Rényi rewiring occurs at constant density ($\langle k \rangle  = 20$) for different values of $l$ and $f$, where $l \times f$ is kept constant to fix the total integration time. Both figures were produced with parameters $m = 2$, $ \alpha = 0.3$ (i.e., with inertia). (a) Here, $l = 2.5 \times 10^4$ and $f = 100$. (b) Here, $l = 5 \times 10^3$ and $f = 500$. All curves depict averages over 25 instantiations.}
  \label{fig:11}
\end{figure*}

\begin{figure*}[t]
\begin{subfigure}[t]{0.02\textwidth}
  \vspace{-6cm}
  (a)
\end{subfigure}
  \adjustbox{minipage=1.3em,valign=t}{\label{sfig:testa}}%
  \begin{subfigure}[t]{\dimexpr.485\linewidth-1.3em\relax}
  \centering
      \hspace*{.8cm}\includegraphics[trim = 1350 0 1000 0, scale=.15]{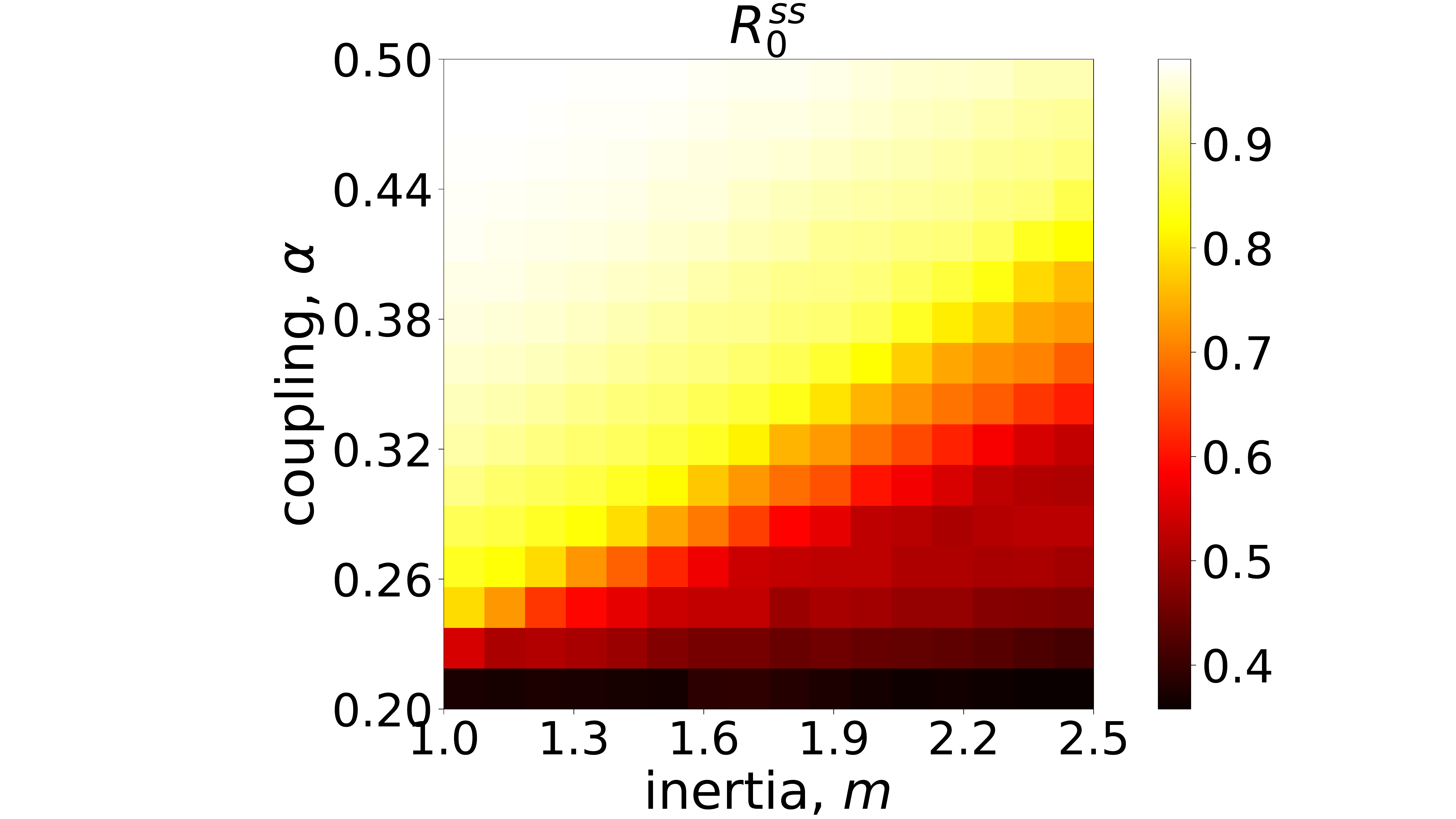}
  \end{subfigure}%
  \begin{subfigure}[t]{0.02\textwidth}
  \vspace{-6cm}
  (b)
\end{subfigure}
  \adjustbox{minipage=1.3em,valign=t}{\label{sfig:testb}}%
  \begin{subfigure}[t]{\dimexpr.485\linewidth-1.3em\relax}
  \centering
  \hspace*{.8cm}\includegraphics[trim = 1350 0 1000 0, scale=.15]{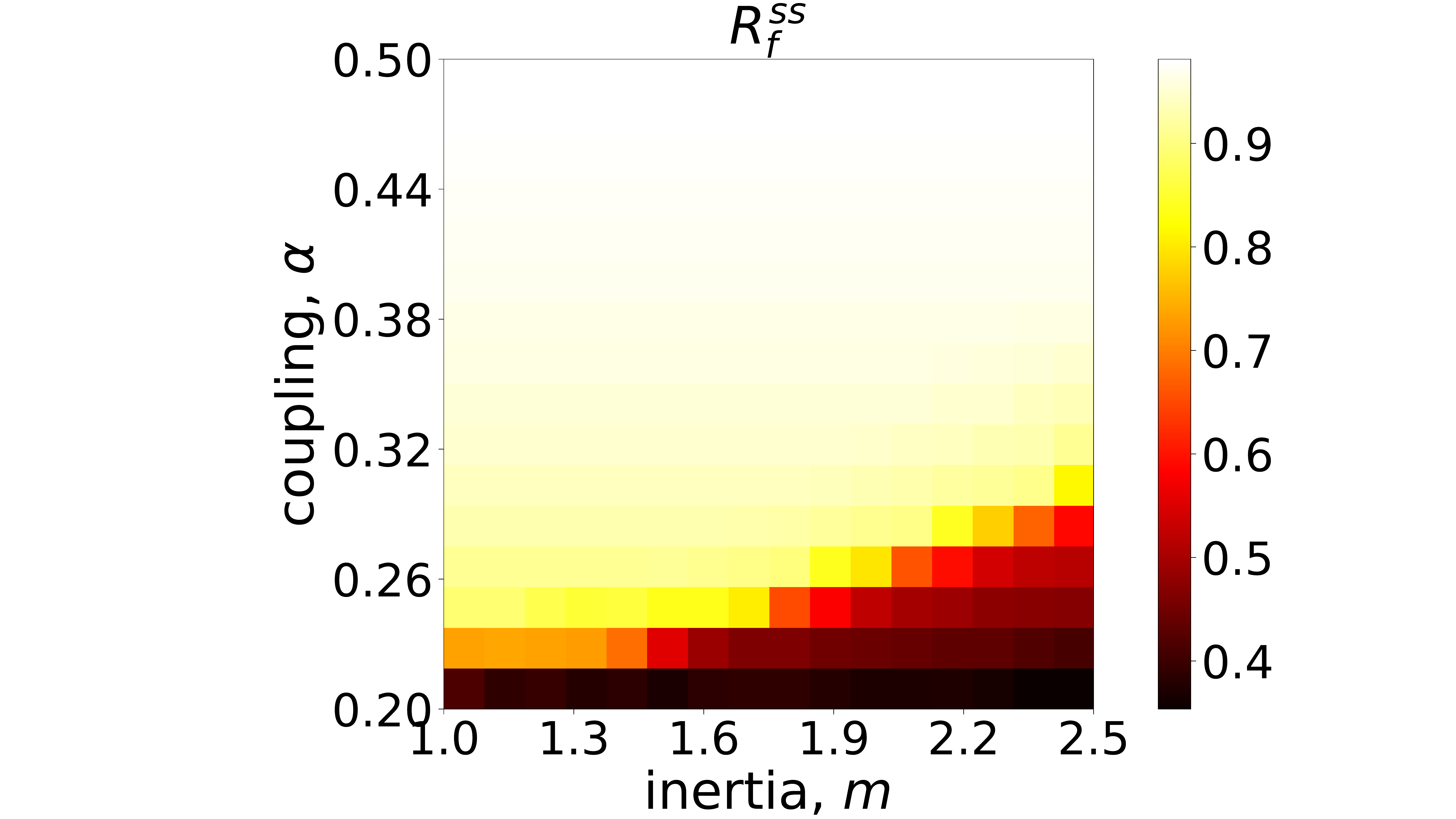}
  \end{subfigure}
      \caption{\textbf{Supplementary analysis of constant-density rewiring over the inertia-coupling parameter space.} \emph{(a)} The initial steady-state order parameter $R_0^{ss}$ prior to Erdős–Rényi $\rightarrow$ synchrony-aligned $\rightarrow$ Erdős–Rényi rewiring. \emph{(b)} The final steady-state order parameter $R_f^{ss}$ after Erdős–Rényi $\rightarrow$ synchrony-aligned $\rightarrow$ Erdős–Rényi rewiring.}
  \label{fig:12}
\end{figure*}

\begin{figure*}[t]
\begin{subfigure}[t]{0.02\textwidth}
  \vspace{-6cm}
  (a)
\end{subfigure}
  \adjustbox{minipage=1.3em,valign=t}{\label{sfig:testa}}%
  \begin{subfigure}[t]{\dimexpr.485\linewidth-1.3em\relax}
  \centering
  \hspace*{.8cm}\includegraphics[trim = 80 0 0 0, scale=.5]{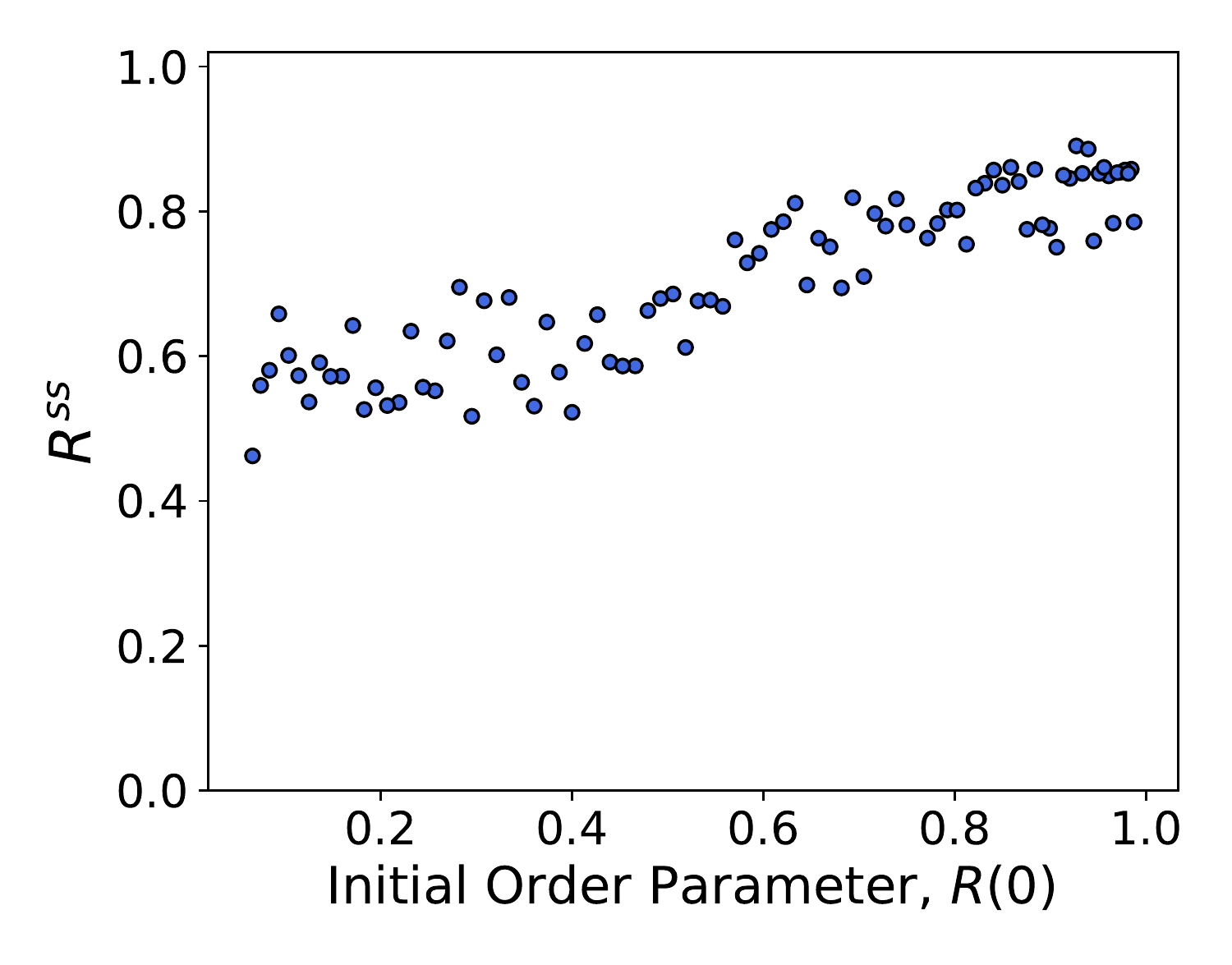}
  \end{subfigure}%
  \begin{subfigure}[t]{0.02\textwidth}
  \vspace{-6cm}
  (b)
\end{subfigure}
  \adjustbox{minipage=1.3em,valign=t}{\label{sfig:testb}}%
  \begin{subfigure}[t]{\dimexpr.485\linewidth-1.3em\relax}
  \centering
  \hspace*{.8cm}\includegraphics[trim = 80 0 0 0, scale=.5]{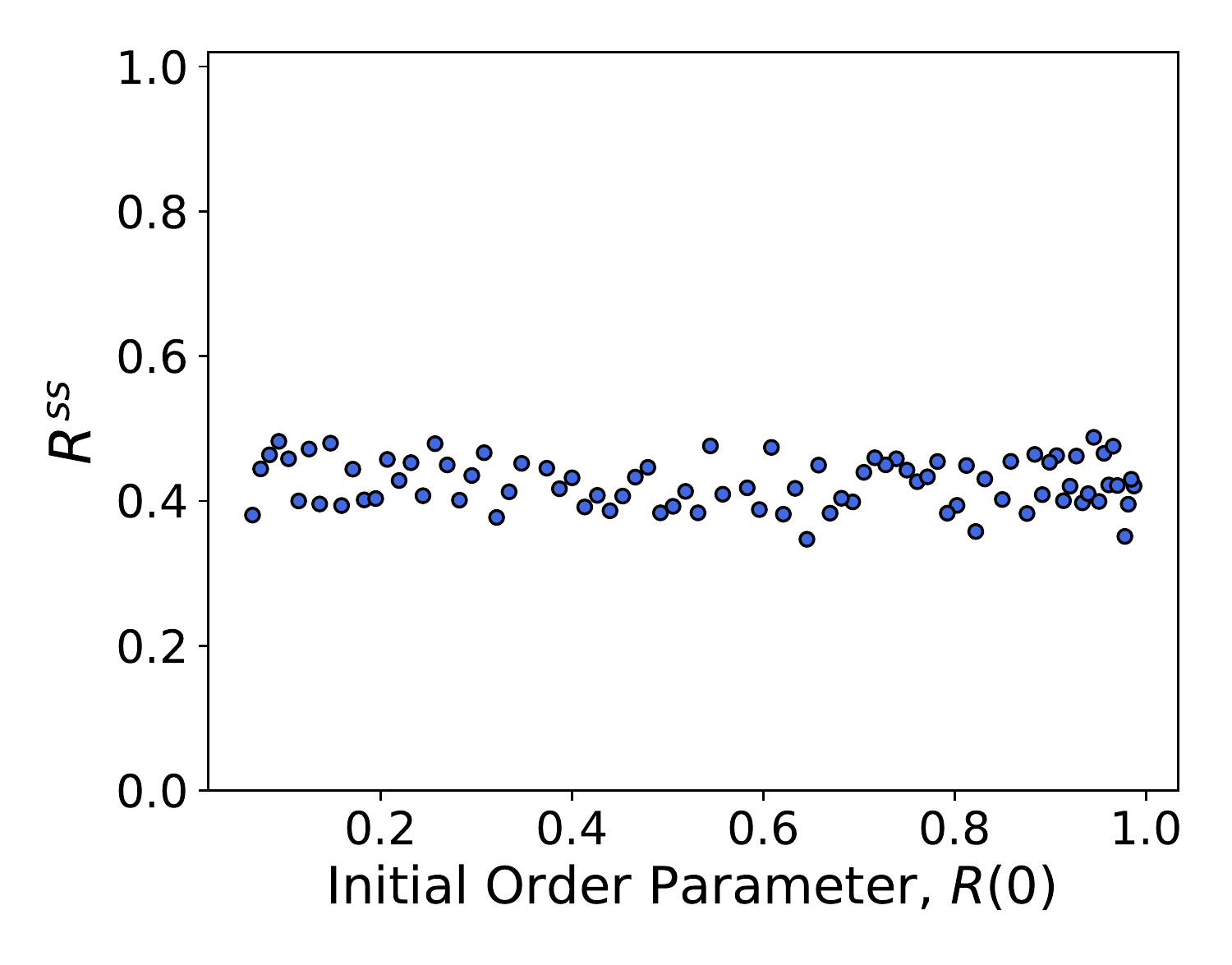}
  \end{subfigure}
      \caption{\textbf{The effect of the initial order parameter on networks of inertial oscillators with static Erdős–Rényi connectivity.} The steady-state order parameter $R^{ss}$ resulting from static Erdős–Rényi connectivity for $100$ different sets of initial phases, with parameters \emph{(a)} $\alpha = 0.3$ and $m = 2$, and parameters \emph{(b)} $\alpha = 0.24$ and $m = 2.3$, corresponding to the parameters of Figs. 5b and 6d, respectively. Initial phases were sampled so that various levels of initial synchrony are represented.}
  \label{fig:13}
\end{figure*}

\begin{figure*}[t]
\begin{subfigure}[t]{0.02\textwidth}
  \vspace{-6cm}
  (a)
\end{subfigure}
  \adjustbox{minipage=1.3em,valign=t}{\label{sfig:testa}}%
  \begin{subfigure}[t]{\dimexpr.485\linewidth-1.3em\relax}
  \centering
  \hspace*{.8cm}\includegraphics[trim = 80 0 0 0, scale=.5]{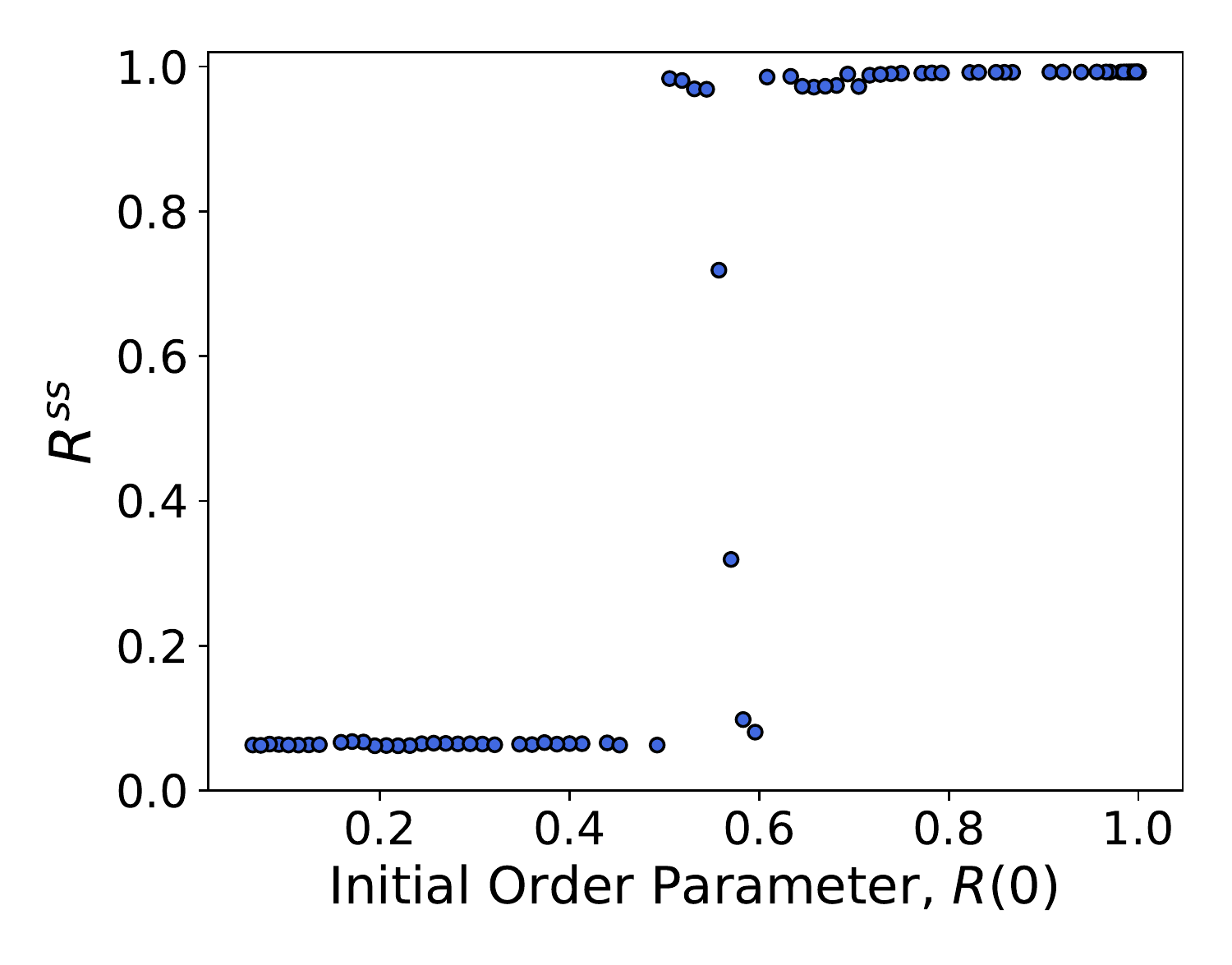}
  \end{subfigure}%
  \begin{subfigure}[t]{0.02\textwidth}
  \vspace{-6cm}
  (b)
\end{subfigure}
  \adjustbox{minipage=1.3em,valign=t}{\label{sfig:testb}}%
  \begin{subfigure}[t]{\dimexpr.485\linewidth-1.3em\relax}
  \centering
  \hspace*{.8cm}\includegraphics[trim = 80 0 0 0, scale=.5]{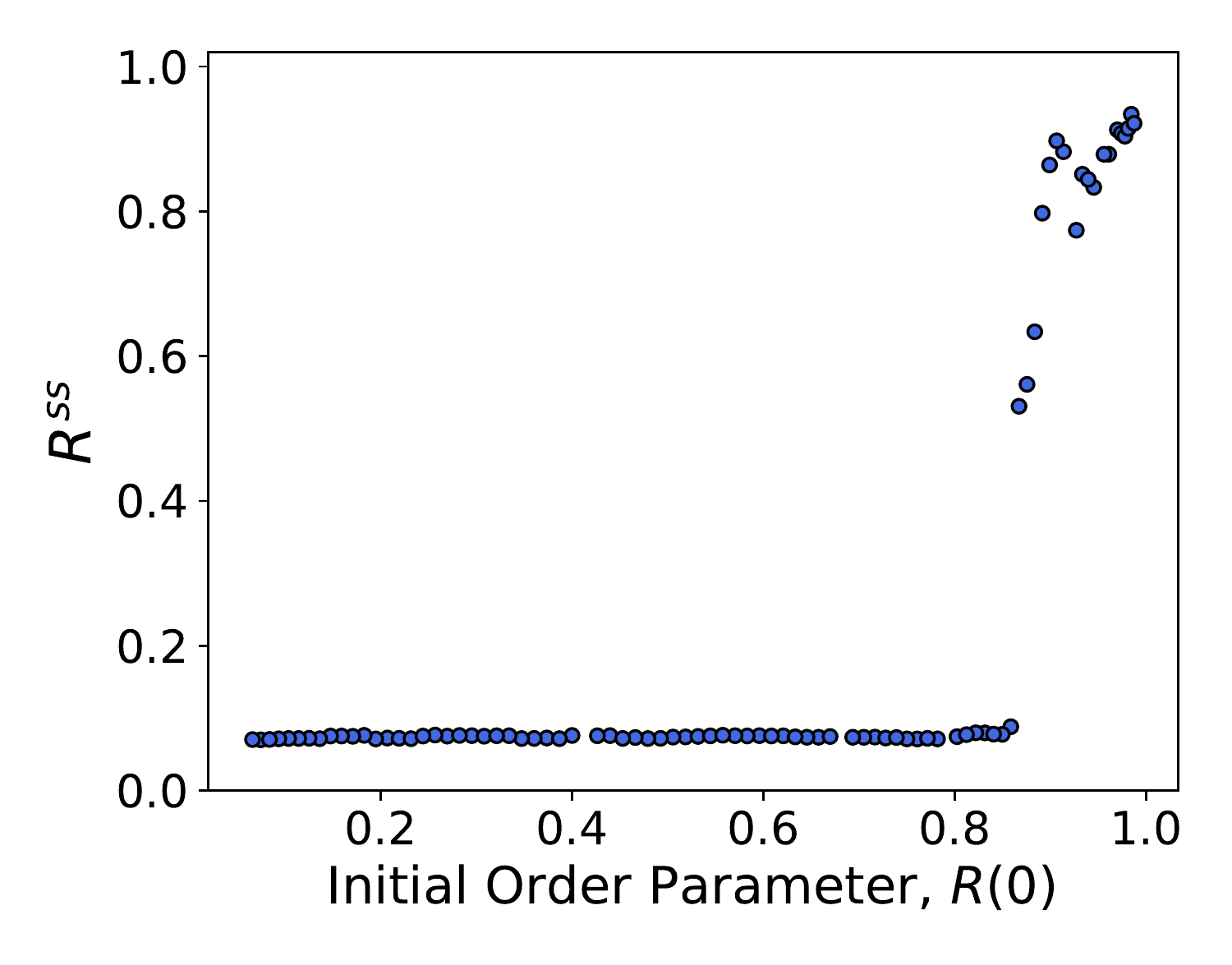}
  \end{subfigure}
      \caption{\textbf{The effect of the initial order parameter on networks of inertial oscillators with static synchrony-aligned connectivity.} The steady-state order parameter $R^{ss}$ resulting from static synchrony-aligned connectivity for $100$ different sets of initial phases, with parameters \emph{(a)} $\alpha = 0.3$ and $m = 2$, and parameters \emph{(b)} $\alpha = 0.24$ and $m = 2.3$, corresponding to the parameters of Figs. 5b and 6d, respectively. Initial phases were sampled so that various levels of initial synchrony are represented.}
  \label{fig:14}
\end{figure*}

\begin{figure*}[t!]
\begin{subfigure}[t]{0.02\textwidth}
  \vspace{-5.5cm}
  (a)
\end{subfigure}
  \adjustbox{minipage=1.3em,valign=t}{\label{sfig:testa}}%
  \begin{subfigure}[t]{\dimexpr.485\linewidth-1.3em\relax}
  \centering
      \includegraphics[trim = 0 0 0 000, scale=.45]{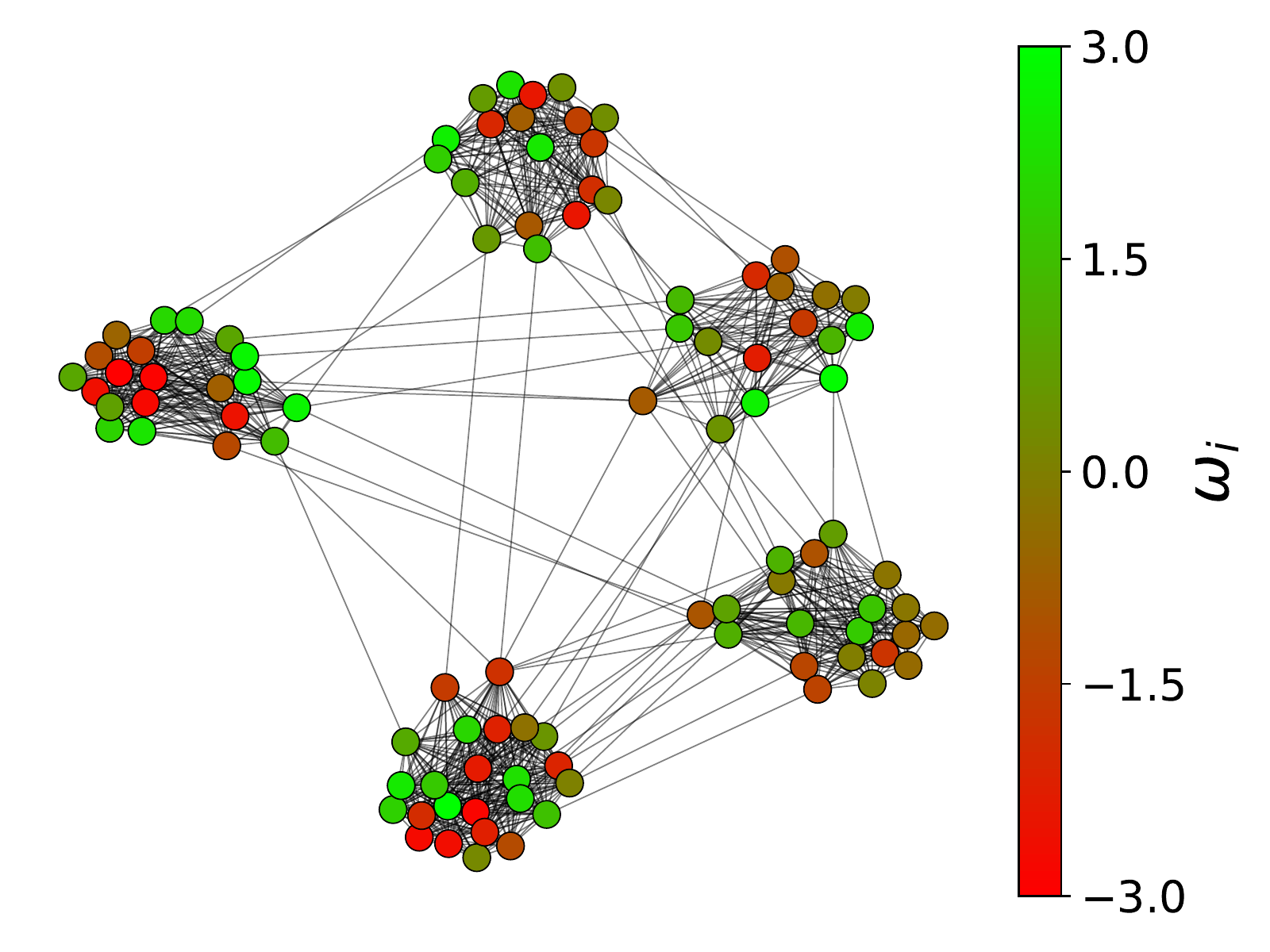}
  \end{subfigure}%
  \begin{subfigure}[t]{0.02\textwidth}
  \vspace{-5.5cm}
  (c)
\end{subfigure}
  \adjustbox{minipage=1.3em,valign=t}{\label{sfig:testb}}%
  \begin{subfigure}[t]{\dimexpr.485\linewidth-1.3em\relax}
  \centering
  \includegraphics[trim = 0 0 0 000, scale=.17]{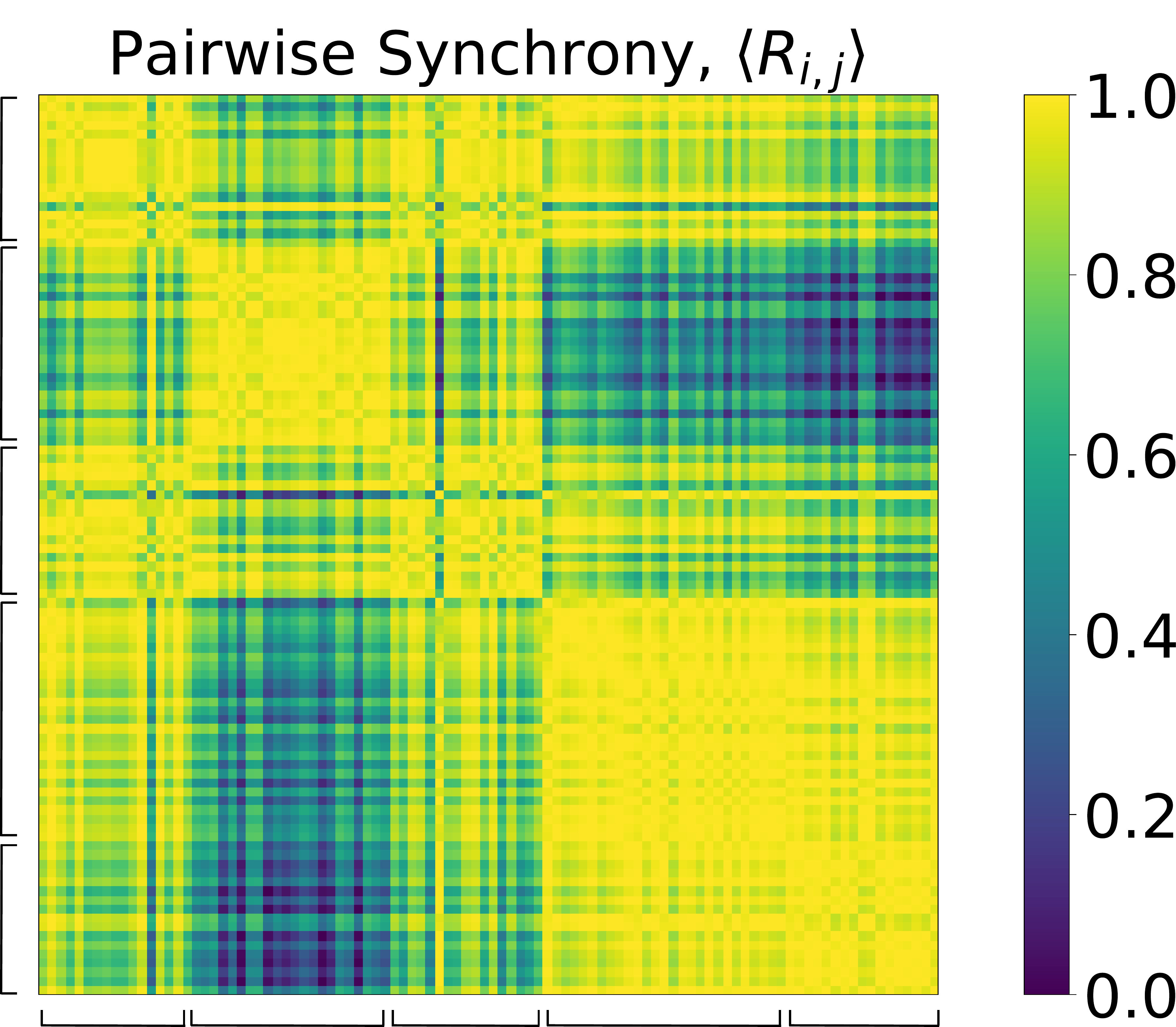}
  \end{subfigure}
  \\
  \begin{subfigure}[t]{0.02\textwidth}
  \vspace{0.5cm}
  (b)
\end{subfigure}
  \adjustbox{minipage=1.3em,valign=t}{\label{sfig:testa}}%
  \begin{subfigure}[t]{\dimexpr.485\linewidth-1.3em\relax}
  \vspace{1em}
  \centering
       \includegraphics[trim = 0 0 0 0, scale=.45]{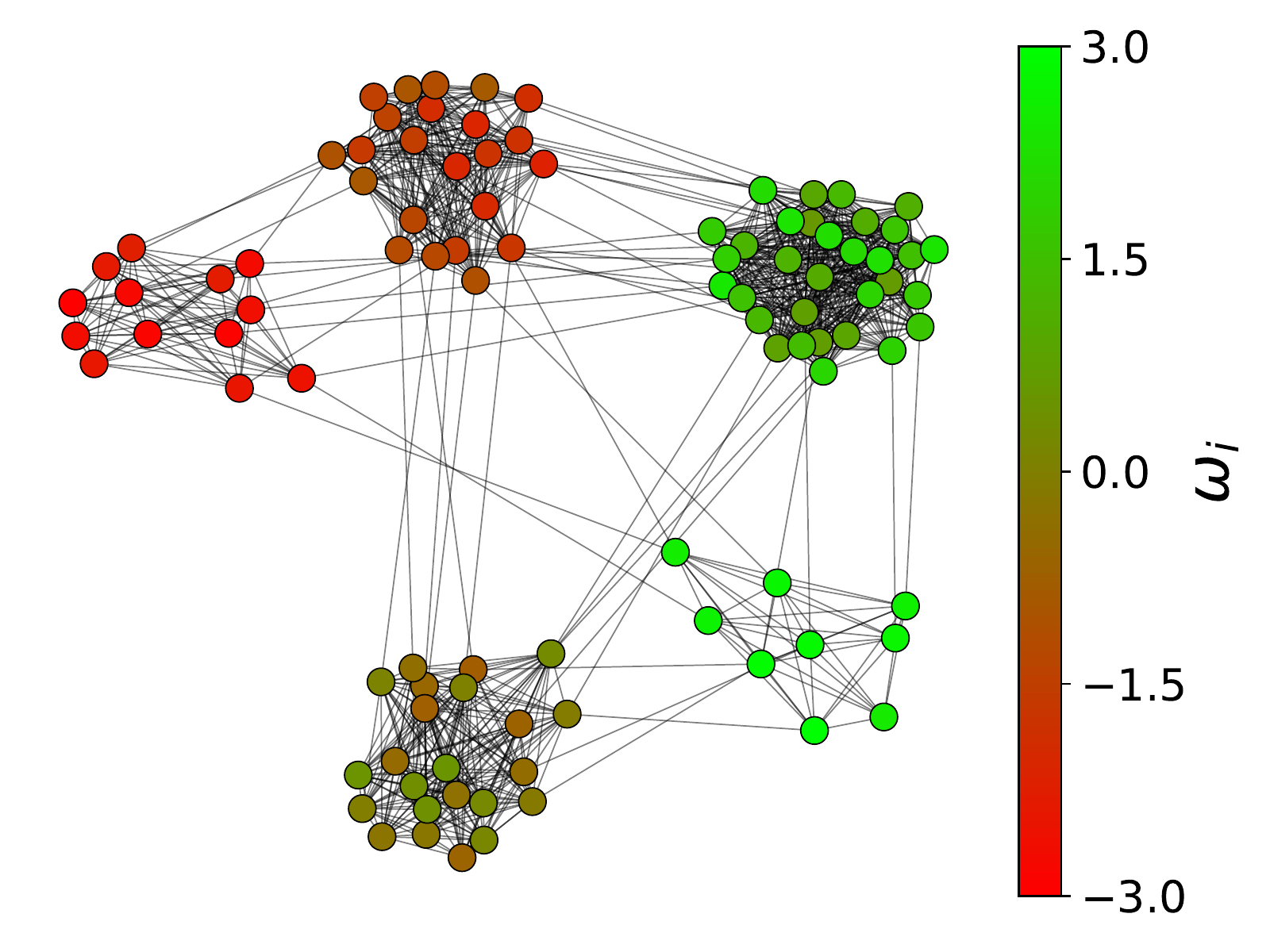}
  \end{subfigure}%
  \begin{subfigure}[t]{0.02\textwidth}
  \vspace{0.5cm}
  (d)
\end{subfigure}
  \adjustbox{minipage=1.3em,valign=t}{\label{sfig:testb}}%
  \begin{subfigure}[t]{\dimexpr.485\linewidth-1.3em\relax}
  \vspace{1em}
  \centering
  \includegraphics[trim = 0 0 0 0, scale=.17]{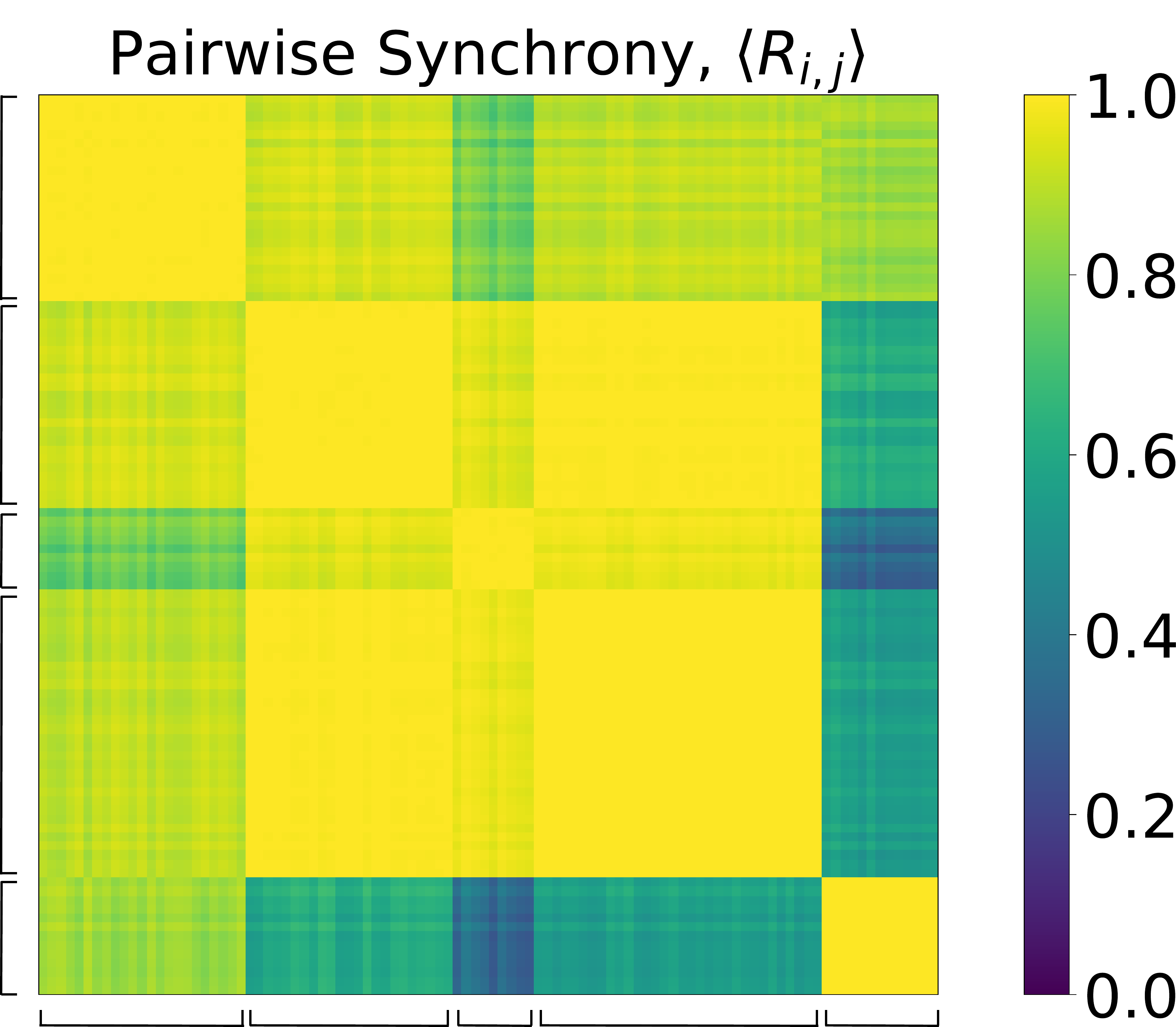}
  \end{subfigure}
  \vspace{.5em}
      \caption{\textbf{Modular networks and their synchronizability during network evolution.} Here we show examples of modular networks used for $G_f$ in Fig.~\ref{fig:7}b. \emph{(a, b)} Examples of a random modular network and a frequency modular network used. Natural frequencies of the oscillators are shown. \emph{(c,d)} The pairwise synchrony $\langle R_{i,j} \rangle$ corresponding to these examples is shown for all pairs of vertices while the $G_f$ topology (\emph{(c)} random modular, or \emph{(d)}, frequency modular) is present during Erdős–Rényi $\rightarrow G_f \rightarrow$ Erdős–Rényi network evolution. Rows and columns are sorted by modules, with modules delimited on the axes. Initial conditions were obtained from the high-synchrony states resulting from Erdős–Rényi $\rightarrow$ synchrony-aligned $\rightarrow$ Erdős–Rényi evolution (Fig.~\ref{fig:5}b).}
  \label{fig:15}
\end{figure*}

\clearpage 

\bibliography{refs}
\bibliographystyle{unsrt}

\end{document}